\documentclass[fleqn,usenatbib]{mnras}

\usepackage{newtxtext,newtxmath}
\usepackage[T1]{fontenc}

\DeclareRobustCommand{\VAN}[3]{#2}
\let\VANthebibliography\thebibliography
\def\thebibliography{\DeclareRobustCommand{\VAN}[3]{##3}\VANthebibliography}

\usepackage{graphicx}	% Including figure files
\usepackage{amsmath}	% Advanced maths commands

\usepackage{booktabs}
\usepackage{threeparttable}
\usepackage{tabularx}

\usepackage{mathtools, nccmath}

\title[SFR variability from EBL and galaxy surveys]{Probing bursty star formation by cross-correlating extragalactic background light and galaxy surveys}

\author[G. Sun et al.]{
Guochao Sun,$^{1}$\thanks{E-mail: guochao.sun@northwestern.edu}
Adam Lidz,$^{2}$
Andreas L. Faisst$^{2,3}$
and Claude-Andr\'{e} Faucher-Gigu\`{e}re$^{1}$
\\
$^{1}$CIERA and Department of Physics and Astronomy, Northwestern University, 1800 Sherman Ave, Evanston, IL 60201, USA \\
$^{2}$University of Pennsylvania, Department of Physics \& Astronomy, 209 S. 33rd Street, Philadelphia, PA 19104, USA \\
$^{3}$Caltech/IPAC, MS314-6, 1200 E. California Boulevard, Pasadena, CA 91125, USA
}

\date{Accepted XXX. Received YYY; in original form ZZZ}

\pubyear{2023}

\begin{document}
\label{firstpage}
\pagerange{\pageref{firstpage}--\pageref{lastpage}}
\maketitle

\begin{abstract}
Understanding the star formation rate (SFR) variability and how it depends on physical properties of galaxies is important for developing and testing the theory of galaxy formation. We investigate how statistical measurements of the extragalactic background light (EBL) can shed light on this topic and complement traditional methods based on observations of individual galaxies. Using semi-empirical models of galaxy evolution and SFR indicators sensitive to different star formation timescales (e.g., H$\alpha$ and UV continuum luminosities), we show that the SFR variability, quantified by the joint probability distribution of the SFR indicators (i.e., the bivariate conditional luminosity function), can be characterized as a function of galaxy mass and redshift through the cross-correlation between deep, near-infrared maps of the EBL and galaxy distributions. As an example, we consider combining upcoming SPHEREx maps of the EBL with galaxy samples from Rubin/LSST. We demonstrate that their cross-correlation over a sky fraction of $f_\mathrm{sky}\sim0.5$ can constrain the joint SFR indicator distribution at high significance up to $z\sim2.5$ for mass-complete samples of galaxies down to $M_{*}\sim10^9\,M_{\odot}$. These constraints not only allow models of different SFR variability to be distinguished, but also provide unique opportunities to investigate physical mechanisms that require large number statistics such as environmental effects. The cross-correlations investigated illustrate the power of combining cosmological surveys to extract information inaccessible from each data set alone, while the large galaxy populations probed capture ensemble-averaged properties beyond the reach of targeted observations towards individual galaxies. 
\end{abstract}


%%%%%%%%%%%%%%%%%%%%%%%%%%%%%%%%%%%%%%%%%%%%%%%%%%

%%%%%%%%%%%%%%%%% BODY OF PAPER %%%%%%%%%%%%%%%%%%

\section{Introduction}

Both observations of star-forming galaxies at different cosmic epochs \citep{Weisz_2012, Emami_2019, Faisst_2019} and galaxy simulations resolving the gravitational collapse of star-forming gas and stellar feedback \citep{Dominguez_2015, Sparre_2017, Gurvich_2023, Hopkins_2023} have led to an emerging picture where the star formation rate (SFR) of galaxies in certain regimes is highly time-variable --- a situation often referred to as bursty star formation. Elucidating the physical origin of bursty star formation and the transition to time-steady star formation is a key task for galaxy formation theory \citep{CAFG_2018, Caplar_2019, Iyer_2020, FM_2022, Orr_2022, Hopkins_2023}. To this end, a crucial way to connect observations with theory is to investigate the variety of SFR indicators sensitive to different timescales of star formation. Among the large number of SFR indicators proposed in the literature, the H$\alpha$ $\lambda$6563 nebular line emission and the UV continuum emission are most commonly considered \cite[e.g.,][]{Emami_2019,JFV_2021}. Because H$\alpha$ emission is predominantly produced by recombinations in \ion{H}{ii} regions ionized by young, massive stars, it is expected to be sensitive to recent SFR variations on timescales as short as a few Myr. On the other hand, the UV continuum emission has substantial contributions from the non-ionizing radiation of older stellar populations and therefore is sensitive to significantly longer star formation timescales ($\sim$10\,Myr when the SFR is time-steady and $\sim$100\,Myr following extreme starbursts; see e.g., \citealt{JFV_2021}). The exact value depends on various factors, such as the wavelength of emission, the star formation history (SFH), and the stellar population synthesis (SPS) model assumed. 

Traditional methods relying on these SFR indicators usually require measuring the H$\alpha$ and UV luminosities of individual galaxies simultaneously from flux limited surveys. Such measurements are expensive and likely susceptible to issues like selection bias that preferentially selects galaxies experiencing an ongoing burst of star formation \citep{Dominguez_2015,Faisst_2019,Sun_2023}. Meanwhile, measuring the mean ratio $L_\mathrm{H\alpha}/L_\mathrm{UV}$ (where $L_\mathrm{UV} = \nu L_\mathrm{\nu}$ is the UV luminosity per logarithmic frequency) alone for a limited sample of galaxies is insufficient to probe the SFR variability because it can be very sensitive to complications such as dust attenuation, whereas characterizing the joint distribution of $L_\mathrm{H\alpha}$ and $L_\mathrm{UV}$, especially its width, with a large galaxy sample can be a lot more informative \citep{Sparre_2017, Emami_2019}. These limitations together make an extensive, mass-complete study of bursty star formation in galaxies of different properties at different cosmic times challenging. 

Composed of the accumulated radiation from the all the sources in the universe outside the Milky Way, the extragalactic background light (EBL) offers a wealth of information about the galaxy and star formation physics across cosmic time \citep{Finke_2010, Finke_2022}. At near-infrared wavelengths (corresponding to rest-frame optical/UV at high redshifts), its potential to constrain the star formation process in high-redshift galaxies have attracted increasing interest in recent years \cite[see e.g.,][]{Sun_2021, Sun_2022, Mirocha_2022, Scott_2022}. Therefore, as an alternative approach to probe bursty star formation, we investigate in this work the possibility of statistically constraining the joint distribution of $L_\mathrm{H\alpha}$ and $L_\mathrm{UV}$ by cross-correlating cosmological surveys of the near-infrared EBL and galaxy distributions. Thanks to its unprecedented survey depth and sky coverage, the SPHEREx mission \citep{Dore_2014,Korngut_2018,Crill_2020} promises to accurately quantify sources of the EBL out to the epoch of reionization and thereby probe galaxy formation and evolution across a wide range of cosmic times. In synergy with wide-field galaxy surveys to be conducted by e.g., the Rubin Observatory Legacy Survey of Space and Time (Rubin/LSST; \citealt{LSST_2009}) or the Nancy Grace Roman Space Telescope \citep{Spergel_2015}, it has been demonstrated that the EBL--galaxy cross-correlation can be detected at high significance in each spectral channel of SPHEREx, thereby allowing the mean rest-frame optical/UV emission spectrum of galaxies to be accurately measured \citep{CC_2022}. It is therefore interesting to explore whether the EBL--galaxy cross-correlation can help constrain bursty star formation in galaxies, including its mass and redshift dependence, and provide a test of galaxy formation theory. 

\begin{figure*}
	\includegraphics[width=\textwidth]{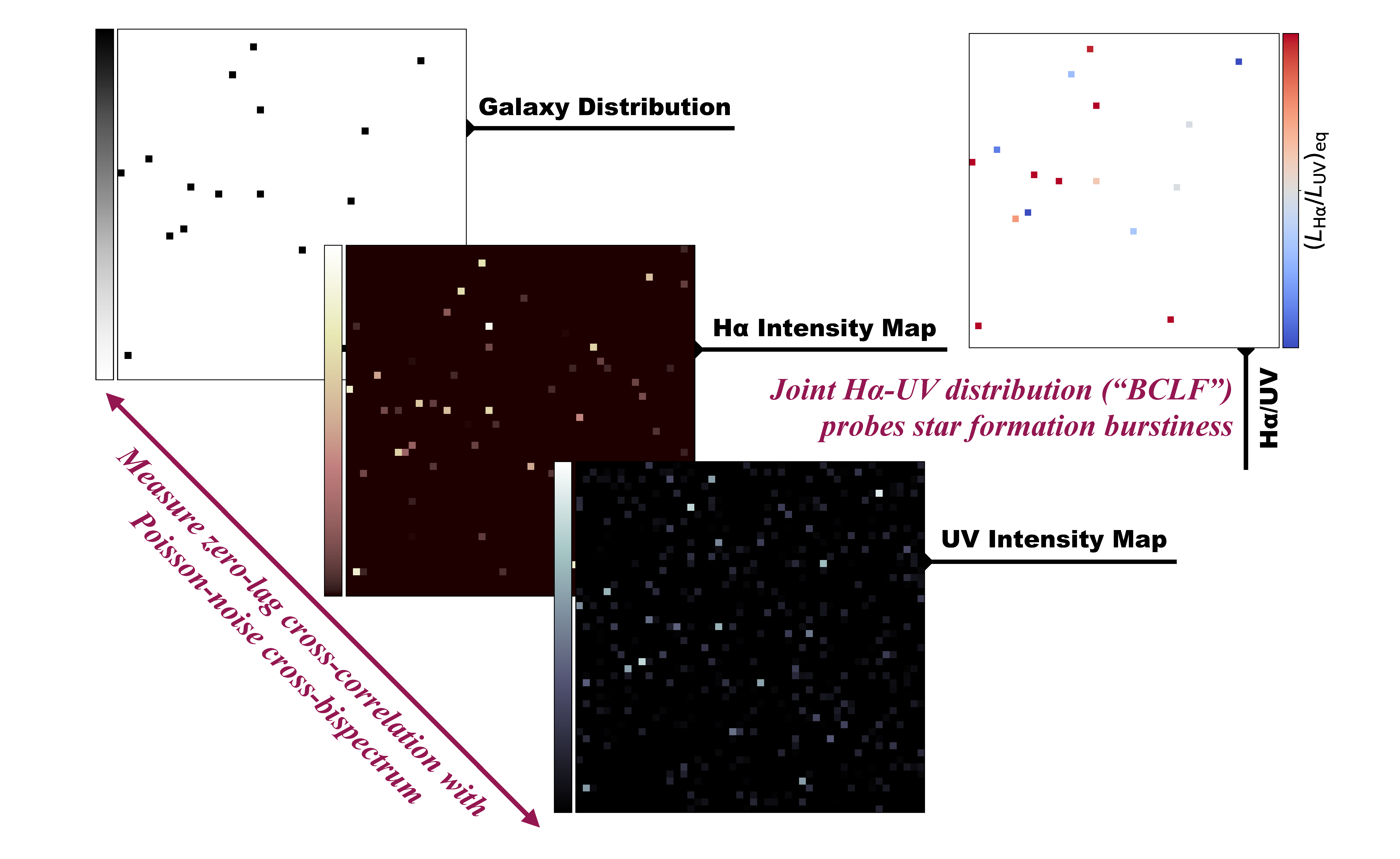}
     \caption{A graphical representation of the EBL--galaxy cross-correlation analysis investigated in this work for probing effects of bursty star formation on the joint distribution of SFR indicators $L_\mathrm{H\alpha}$ and $L_\mathrm{UV}$. Distributions of galaxies, H$\alpha$ line intensity, and UV continuum intensity are cross-correlated in Fourier space to measure the cross-bispectrum. This constrains the joint H$\alpha$--UV luminosity distribution, especially its width which reflects the scatter in $L_\mathrm{H\alpha}/L_\mathrm{UV}$ around the equilibrium value when star formation is time-steady. The Fourier-space cross-bispectrum analysis in the Poisson-noise dominated limit is formally equivalent to a zero-lag cross-correlation (i.e., stacking) on galaxy positions in real space, as demonstrated in Section~\ref{sec:realvsfourier}, but allows foregrounds and observational systematics to be more easily separated (Section~\ref{sec:discussion}).}
    \label{fig:visualization}
\end{figure*}

In this paper, we conduct a proof-of-principle study of using the (near-infrared) EBL--galaxy cross-correlation to probe bursty star formation. In particular, we focus on the cross-correlation between intensity maps of H$\alpha$ and UV continuum emission and the distribution of galaxies selected by their stellar mass. More specifically, we aim to constrain the joint distribution of $L_\mathrm{H\alpha}$ and $L_\mathrm{UV}$ as a probe for the SFR variability by measuring the zero-lag cross-correlation of the distribution of mass-selected galaxy samples and intensity maps of H$\alpha$ and UV emission. As illustrated in Fig.~\ref{fig:visualization}, such a measurement can probe the decorrelation effect on the zero-lag cross-correlation caused by the scatter in the $L_\mathrm{H\alpha}$--$L_\mathrm{UV}$ joint distribution, which links to the SFR variability (though complications due to e.g., dust attenuation exist; see Section~\ref{sec:discussion}). To measure the zero-lag cross-correlation, we calculate the Poisson-noise cross-bispectrum in Fourier space, which is the optimal way to separate the signal of interest from other sources of confusion, including large-scale clustering, instrument noise, and observational systematics. We forecast the prospects for measuring this cross-correlation using SPHEREx and Rubin/LSST and demonstrate the utility for probing bursty star formation in galaxies in different mass and redshift ranges. 

% Figure 1

The remainder of this paper is organized as follows. In Section~\ref{sec:methods}, we first introduce a simple, semi-empirical model for the $L_\mathrm{H\alpha}$--$L_\mathrm{UV}$ joint distribution of galaxies conditioned on stellar mass. We then show, in the limit where Poisson fluctuations dominate over clustering, how the zero-lag cross-correlation in real space is equivalent to a measurement of the cross-bispectrum in Fourier space. Finally, we describe the full framework for constraining the $L_\mathrm{H\alpha}$--$L_\mathrm{UV}$ joint distribution with a set of correlation coefficients defined by cross-bispectra. In Section~\ref{sec:results}, we present main results of our analysis, including forecasts for the various cross-correlation signals and the implied constraints on the toy models considered in our case study for SPHEREx and Rubin/LSST. We discuss some limitations and caveats of the presented analysis in Section~\ref{sec:discussion}, before concluding in Section~\ref{sec:conclusions}. A flat, $\Lambda$CDM cosmology consistent with the measurements from \citet{Planck_2016XIII} is adopted throughout this paper. 

\section{Methods} \label{sec:methods}

\subsection{Modeling the $L_\mathrm{H\alpha}$--$L_\mathrm{UV}$ joint distribution}

\subsubsection{Overview} \label{sec:model_overview}

While the modeling and analysis frameworks to be presented are generally applicable, for our proof-of-principle study in this paper, we investigate specifically the prospects for cross-correlating near-infrared EBL maps measured by SPHEREx with distributions of galaxies from the Rubin/LSST photometric redshift survey, which is expected to measure the mean rest-UV/optical spectrum of galaxies at high significance up to $z\sim4$ \citep{CC_2022}. 

Given the wavelength coverage of SPHEREx (0.75--5$\mu$m) and the redshift range over which high-quality photo-$z$ measurements can be achieved by Rubin/LSST, we aim to optimize the chance of detecting the decorrelation between H$\alpha$ and UV luminosities due to bursty star formation, which is expected to be more pronounced in low-mass galaxies that are abundant but faint. For the longer-timescale SFR indicator, we choose the $U$-band (3500\,\AA) luminosity\footnote{Throughout this paper, we use UV and $U$-band interchangeably when referring to the continuum emission to be studied together with H$\alpha$. For simplicity, we refer to it with the subscript $U$ hereafter.} rather than the more commonly used FUV (1500\,\AA) luminosity because the former reaches lower redshifts ($z\simeq1.2$) and maximizes the contrast in star formation timescales compared to H$\alpha$ \citep{Emami_2021}. 

Performing the analysis at $z\sim1$ rather than $z\sim4$, is also motivated by the completeness limit of the Rubin/LSST photometric redshift survey, below which issues like selection bias due to incompleteness introduce significant systematics. Following \citet{Leauthaud_2020}, we can estimate the stellar mass range accessible by scaling from the 90\% mass completeness limit of the COSMOS2015 catalog \citep{Laigle_2016}. For Rubin/LSST with $i$-band limiting magnitude of $i=26.8$, the 90\% mass completeness limits are $\log(M^\mathrm{lim}_{*}/M_{\odot}) = 8.55, 8.95, 9.25, 9.4$ at $z=1, 1.5, 2, 2.5$, respectively, well below stellar masses at which simulations predict galaxies at these redshifts to exhibit a considerable level of scatter in $L_\mathrm{H\alpha}/L_\mathrm{UV}$ due to bursty star formation \citep{Dominguez_2015, Sparre_2017}. 

We analytically derive a conditional luminosity function (CLF)-based description of the different moments of $L_\mathrm{H\alpha}$ and $L_{U}$ necessary for the cross-correlation. Since galaxies in different stellar mass bins will be analyzed separately, the luminosity distributions are conditioned on stellar mass $M_{*}$. The exact parameterization is based on semi-empirical models of galaxy evolution and H$\alpha$ and UV emission, which are verified against the matching between the observed $U$-band luminosity functions \cite[e.g.,][]{Moutard_2020} and stellar mass functions \cite[e.g.,][]{Shuntov_2022} at redshifts of interest. 

\subsubsection{H$\alpha$--UV bivariate conditional luminosity function (BCLF)}

Taking $\Phi(L)$ to be the probability distribution function (PDF) of the luminosity $L$ such that $\int \Phi(L) d L = 1$, we can write the joint PDF of $L_\mathrm{H\alpha}$ and $L_{U}$ conditioned on $M_{*}$ as
\begin{equation}
\Phi(L_\mathrm{H\alpha}, L_{U}|M_{*}) = \Phi(L_\mathrm{H\alpha} | L_{U}, M_*) \Phi(L_{U}|M_*),
\label{eq:phi_chain}
\end{equation}
where on the right-hand side the first term $\Phi(L_\mathrm{H\alpha} | L_{U}, M_*)$ is given by a log-normal distribution around the mean H$\alpha$ luminosity $\bar{L}_\mathrm{H\alpha} = L_{U,0}(L_{U}/L_{U,0})^{\beta}$, following the functional form from \citet{Mehta_2015}, with a logarithmic scatter of $\sigma_{\alpha U}(M_{*})$. The second term, $\Phi(L_{U}|M_*)$, often referred to as the \textit{conditional luminosity function} (see e.g., \citealt{Yang_2003}), is the distribution of $L_{U}$ conditioned on $M_{*}$ that can be determined by matching the observed stellar mass function and UV luminosity function. For $\Phi(L_{U}|M_*)$, we also consider a log-normal distribution specified by some mean relation $\bar{L}_{U}(M_{*})$ and a logarithmic scatter $\sigma_\mathrm{LM}$. Putting these ingredients together, we define a \textit{bivariate conditional luminosity function} (BCLF) of $L_\mathrm{H\alpha}$ and $L_{U}$, $\Phi(L_\mathrm{H\alpha}, L_{U} | M_{*})$, that is the product of
\begin{equation}
\Phi(L_\mathrm{H\alpha}|L_{U}, M_{*}) = \frac{\exp\left\{\frac{-[\ln L_\mathrm{H\alpha} - \ln\bar{L}_\mathrm{H\alpha}(L_{U})]^2}{2\sigma_{\alpha U}^2(M_{*})} \right\}}{\sqrt{2\pi}\sigma_{\alpha U}(M_{*}) L_\mathrm{H\alpha}}
\end{equation}
and
\begin{equation}
\Phi(L_{U}|M_*) = \frac{1}{\sqrt{2\pi}\sigma_\mathrm{LM} L_{U}} \exp\left\{\frac{-[\ln L_{U} - \ln\bar{L}_{U}(M_{*})]^2}{2\sigma_\mathrm{LM}^2} \right\},
\label{eq:clf}
\end{equation}
which satisfies
\begin{equation}
\bar{L}_{U}(M_*) = e^{-\sigma^2_\mathrm{LM}/2} \int_{0}^{\infty} d L_{U} \Phi(L_{U}|M_*) L_{U}.
\end{equation}

By the definition of the CLF, equation~(\ref{eq:clf}) can in principle be determined by finding the appropriate functional form of $\bar{L}_{U}(M_{*})$ and the value of $\sigma_\mathrm{LM}$ that best matches the observed $U$-band luminosity function $\phi(L_{U})=dn/dL_{U}$ and stellar mass function $\psi(M_*)=dn/dM_{*}$, where $n$ is the number density of galaxies. In this work, however, we construct a simple, parametric model of $\bar{L}_{U}(M_{*})$ and $\sigma_\mathrm{LM}$ based on the specific SFR--stellar mass relation from semi-empirical models of galaxy formation given by the UniverseMachine code \citep{Behroozi_2019} and the observed $U$-band luminosities of galaxies from \citet{Zhou_2017}. As a sanity check, we have verified our simple model by comparing its predicted $U$-band luminosity function against the observed ones at redshifts where measurements are available \citep{Moutard_2020}. 

To describe H$\alpha$ and $U$-band continuum emission, we take $L_\mathrm{H\alpha} = 2.1 \times 10^{41}\,\mathrm{erg\,s^{-1}} \left(\mathrm{SFR}/\,M_{\odot}\,\mathrm{yr^{-1}}\right)$, valid for the Chabrier IMF \citep{Chabrier_2003} assumed in this work, and adopt the attenuation-corrected, empirical relation between $U$-band and H$\alpha$ luminosities from \citet{Zhou_2017}, who provide a calibration of the $U$-band luminosity as an SFR indicator. Because both these luminosities and the stellar masses they are anchored to are dust-corrected, to properly model their observed strengths in our cross-correlation analysis, we must reapply dust attenuation. To do this self-consistently, we assume the $A_\mathrm{FUV}(M_{*})$ relation from \citet{McLure_2018} that is derived for star-forming galaxies at $z\sim2$--3, 
\begin{equation}
A_\mathrm{FUV} = 2.293 + 1.16\mathcal{M}_{10} + 0.256\mathcal{M}^2_{10} + 0.209\mathcal{M}^3_{10},
\label{eq:a_fuv}
\end{equation}
where $\mathcal{M}_{10}=\log(M_{*}/10^{10}\,M_{\odot})$, and the \citet{Calzetti_2000} dust attenuation curve, which implies $A_\mathrm{H\alpha}=0.44A_\mathrm{FUV}$ and $A_{U}=0.62A_\mathrm{FUV}$, respectively\footnote{Following \citet{McLure_2018}, we assume that $E(B-V)_\mathrm{star} = 0.76 E(B-V)_\mathrm{neb}$ and derive $A_{\lambda}=k_{\lambda} E(B-V)$ from $k_{\lambda}=2.659\times(-2.156+1.509/\lambda-0.198/\lambda^2+0.011/\lambda^3)+4.05$ for $0.12\,\mu\mathrm{m}<\lambda<0.63\,\mu\mathrm{m}$ (rest-frame) or $k_{\lambda}=2.659\times(-1.857+1.040/\lambda)+4.05$ for $0.63\,\mu\mathrm{m}<\lambda<2.2\,\mu\mathrm{m}$, as in \citet{Calzetti_2000}.}. 

\begin{figure}
    \centering
	\includegraphics[width=\columnwidth]{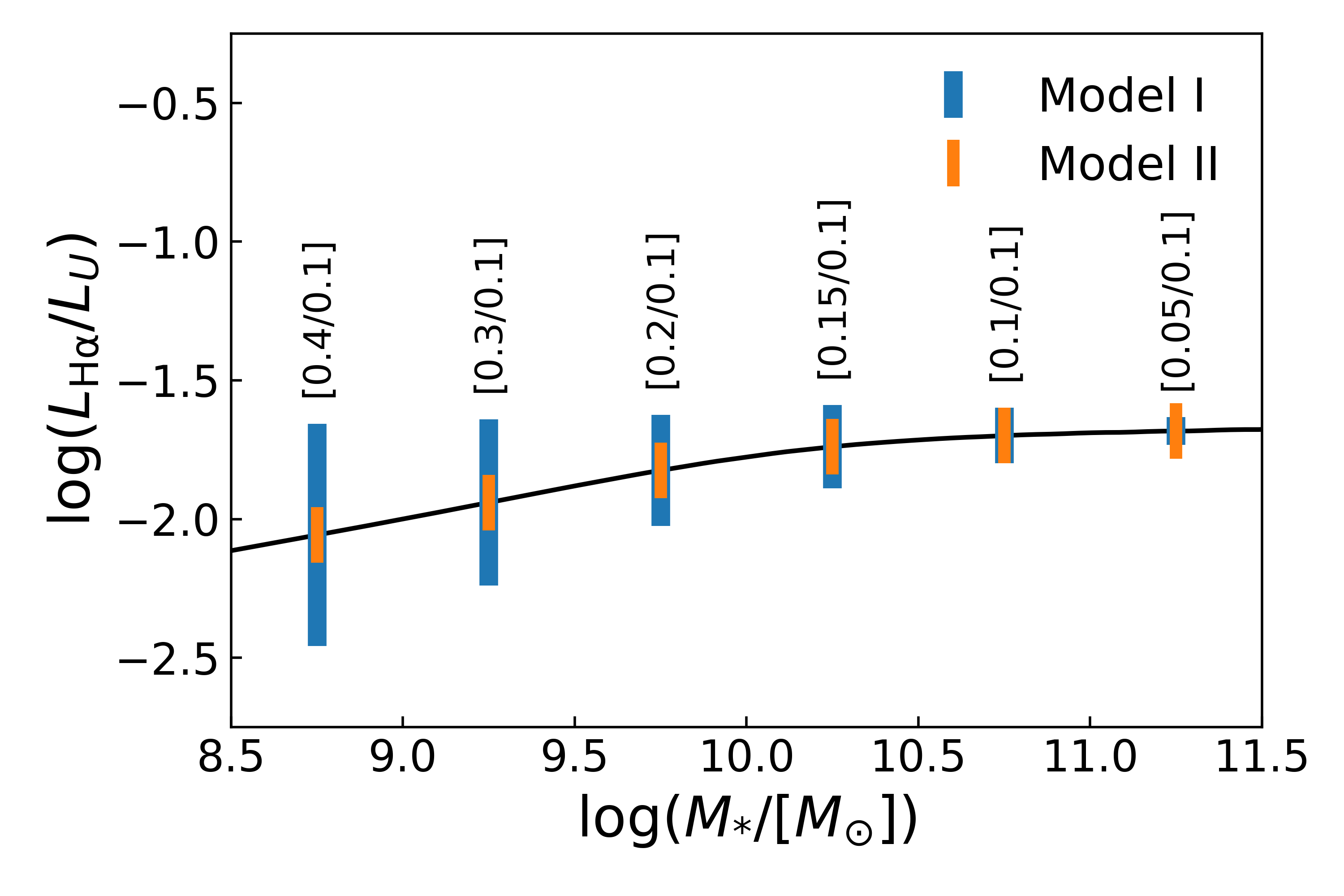}
    \caption{Illustration of the average $\log(L_\mathrm{H\alpha}/L_{U})$ and the scatter around it (specified in brackets in units of dex) as a function of stellar mass described by the baseline model (Model~I) and its variant (Model~II) considered in this work. The scatters are overplotted on the mean relation in the 6 stellar mass bins uniformly distributed over $8.5 < \log(M_{*}/M_{\odot}) < 11.5$. The growth of scatter with decreasing stellar mass as in Model~I is often considered as an indication of an increasing level of bursty star formation.}
    \label{fig:toy_models}
\end{figure}

With the BCLF of $L_\mathrm{H\alpha}$ and $L_{U}$, the ensemble averages that enter our cross-correlation analysis can then be written as
\begin{align}
\langle L_{U} L_\mathrm{H\alpha} \rangle \propto & \int d M_* \Phi(L_\mathrm{H\alpha},L_{U}|M_*) \psi(M_*) \, \times \nonumber \\
& \iint d L_{U} d L_\mathrm{H\alpha} 10^{-0.4(A_\mathrm{H\alpha}+A_{U})} L_{U} L_\mathrm{H\alpha},
\end{align}
\begin{align}
\langle L^2_\mathrm{H\alpha} \rangle \propto & \int d M_* \Phi(L_\mathrm{H\alpha},L_{U}|M_*) \psi(M_*) \, \times \nonumber \\
& \iint d L_{U} d L_\mathrm{H\alpha} 10^{-0.8A_\mathrm{H\alpha}} L^2_\mathrm{H\alpha},
\end{align}
and
\begin{equation}
\langle L^2_{U} \rangle \propto \int d M_* \Phi(L_{U}|M_*) \psi(M_*) \int d L_{U} 10^{-0.8A_{U}} L^2_{U},
\end{equation}
where $\psi(M_{*})$ is the stellar mass function that we self-consistently obtain from UniverseMachine, and $\langle...\rangle$ implicitly assumes that the ensemble average is taken for the sample of stellar-mass-selected galaxies over the mass bin $[M_*, M_*+\Delta M_*]$. We have also confirmed that using the latest observed stellar mass functions \cite[e.g.,][]{Shuntov_2022} has little impact on our results. 

\begin{table}
\centering
\caption{Specifications of the toy models considered in this work. The scatter $\sigma_{\alpha U}$ is allowed to vary across the 6 stellar mass bins uniformly distributed over $8.5 < \log(M_{*}/M_{\odot}) < 11.5$.}
\label{tb:model_params}
\begin{threeparttable}
\begin{tabularx}{\columnwidth}{ccccc}
\toprule
Model & $\sigma_{\alpha U}$ & $\sigma_\mathrm{LM}$ & $\beta$ & ${L_{U,0}}^{\dagger}$ \\
 & (dex) & (dex) & & $(\mathrm{erg\,s^{-1}})$ \\
\hline
I & 0.4, 0.3, 0.2, 0.15, 0.1, 0.05 & 0.2 & 1.25 & $3.55\times10^{51}$ \\
II & 0.1, 0.1, 0.1, 0.1, 0.1, 0.1 & 0.2 & 1.25 & $3.55\times10^{51}$ \\
\bottomrule
\end{tabularx}
\begin{tablenotes}
\item[$\dagger$] The exact value of $L_{U,0}$ does not impact the cross-correlation coefficients (Section~\ref{sec:methods:ccc}) but affects the expected detectability of cross-correlation. 
\end{tablenotes}
\end{threeparttable}
\end{table}

As illustrated in Fig.~\ref{fig:toy_models} and summarized in Table~\ref{tb:model_params}, two toy models of the BCLF of $L_\mathrm{H\alpha}$ and $L_{U}$ are considered for our subsequent analysis. The fiducial model, Model~I, assumes that the scatter, $\sigma_{\alpha U}$, increases with decreasing stellar mass, whereas the contrasting model, Model~II, assumes a constant $\sigma_{\alpha U}=0.1\,$dex across all stellar mass bins. For both models, we further assume a constant $\sigma_\mathrm{LM}=0.2\,$dex, consistent with the scatter in the light-to-mass ratio observed and commonly assumed in semi-empirical models of high-$z$ galaxies \citep{More_2009, SF_2016}, whereas $\beta = 1.25$ and $L_{U,0} = 3.55\times10^{51}\,\mathrm{erg\,s^{-1}}$ are suggested by the best-fit relation to the observed correlation between H$\alpha$ and $U$-band luminosities of galaxies \citep{Zhou_2017}. We note that even though more accurately modeling the H$\alpha$--UV BCLF is beyond the scope of this study, our simple parameterization of the mean relations is grounded on empirical models that reliably describe galaxy evolution and the production of H$\alpha$ and $U$-band emission at the redshifts of interest. The two contrasting cases for $\sigma_{\alpha U}$ are chosen to roughly bracket the range of possible mass dependence of the width of $L_\mathrm{H\alpha}$--$L_{U}$ distribution as a proxy for star formation burstiness, motivated by observations and numerical simulations \citep{Weisz_2012, Dominguez_2015, Sparre_2017, Emami_2019, Faisst_2019}. 

\subsubsection{Connection to the EBL--galaxy cross-correlation} \label{sec:methods:ccc}

From the ensemble averages defined above and their dependence on our BCLF model parameters, we can obtain a few simple and useful expressions that connect cross-correlation observables to these model parameters. The observable most directly related to the cross-correlation analysis is the cross-correlation coefficient, 
$r^\mathrm{g}_\mathrm{\times}(\ell)$, which characterizes how correlated the two SFR tracer fields are for the galaxy population $g$ of interest. As will be shown in Section~\ref{sec:fourierspace}, when measured in the Poisson-noise limit in Fourier space, the cross-correlation coefficient $r^\mathrm{g}_\mathrm{\times,P} \equiv r^\mathrm{g}_\mathrm{\times}(\ell \gg \ell_\mathrm{c})$ takes the simple form
\begin{equation}
r^\mathrm{g}_\mathrm{\times,P} = \frac{B^{U, \mathrm{H\alpha}, \mathrm{g}}_{\ell, \mathrm{P}}}{\sqrt{B^{U, U, \mathrm{g}}_{\ell, \mathrm{P}} B^\mathrm{H\alpha, H\alpha, g}_{\ell, \mathrm{P}}}} \propto \frac{\langle L_{U} L_\mathrm{H\alpha} \rangle}{\sqrt{\langle L_{U}^2 \rangle \langle L_\mathrm{H\alpha}^2 \rangle}}.
\label{eq:r}
\end{equation}
Here, the multipole moment $\ell_\mathrm{c}$ denotes some characteristic scale (to be estimated from a power spectrum analysis) at which non-linear clustering is comparable to the Poisson noise, and $B^{i,j,k}_{\ell,\mathrm{P}}$ denotes the Poisson-noise-limit cross-bispectrum of fields $i$, $j$, and $k$. In Section~\ref{sec:realvsfourier}, we will first motivate the understanding of the cross-correlation of interest in both real and Fourier spaces. We will then detail how to arrive at the proportionality, and derive the components of $r^\mathrm{g}_{\times,\mathrm{P}}$ and their uncertainties, in Section~\ref{sec:fourierspace}. 

Combining equations~(\ref{eq:phi_chain}) through (\ref{eq:r}), we can show that $r^\mathrm{g}_\mathrm{\times,P}$ is in fact insensitive to the $\bar{L}_{U}(M_{*})$ parameterization or the value of $L_{U,0}$, and obtain
\begin{equation}
\ln\left[r^\mathrm{g}_\mathrm{\times,P}\right] = -\left[ \frac{\sigma_{\alpha U}^2}{2} + \frac{\sigma^2_\mathrm{LM} (\beta-1)^2}{2} \right].
\label{eq:lnrx}
\end{equation}
It is easy to see that $r^\mathrm{g}_\mathrm{\times,P}$ drops below unity if either $\sigma_\mathrm{\alpha U}$ or $\sigma_\mathrm{LM}$ (as long as $\beta$ is not strictly 1) is non-zero. While the latter characterizes the intrinsic scatter in the mass-to-light ratio of galaxies due to stochasticity in e.g., mass accretion rates \citep{McBride_2009,Fakhouri_2010,vandenBosch_2014}, the former may be largely attributed to the time variability of the SFR. Because constraints on bursty star formation mainly come from the comparison of $r^\mathrm{g}_\mathrm{\times,P}$ in different stellar mass bins instead of its exact values, factors that are generally mass-independent will not significantly complicate the interpretation. For reference, assuming $\sigma_\mathrm{LM}=0$, we have $r^\mathrm{g}_\mathrm{\times,P} = 0.97$, 0.79, and 0.52 for $\sigma = 0.1$, 0.3, and 0.5\,dex, respectively. 

By analogy to the cross-correlation coefficient, $r^\mathrm{g}_{\times,\mathrm{P}}$, we can also define and derive the following auto-correlation coefficients for H$\alpha$ and UV emission
\begin{equation}
\ln\left(r^\mathrm{g}_\mathrm{H\alpha, P}\right) = \ln \left( \frac{C^\mathrm{H\alpha,g}_{\ell,\mathrm{P}}}{\sqrt{B^\mathrm{H\alpha, H\alpha, g}_{\ell, \mathrm{P}}}} \right) = -\frac{\sigma^2_{\alpha U} + \sigma^2_\mathrm{LM} \beta^2}{2}
\label{eq:lnrha}
\end{equation}
and
\begin{equation}
\ln\left(r^\mathrm{g}_{U,\mathrm{P}}\right) = \ln \left( \frac{C^{U,\mathrm{g}}_{\ell,\mathrm{P}}}{\sqrt{B^{U, U, \mathrm{g}}_{\ell, \mathrm{P}}}} \right) = -\frac{\sigma^2_\mathrm{LM}}{2},
\label{eq:lnruv}
\end{equation}
where $C^\mathrm{H\alpha,g}_{\ell,\mathrm{P}}$ and $C^{U,\mathrm{g}}_{\ell,\mathrm{P}}$ are the Poisson-noise terms of the angular cross-power spectra of H$\alpha$ and UV emission with galaxies to be defined in Section~\ref{sec:fourierspace}. 

Equations~(\ref{eq:lnrx}) through (\ref{eq:lnruv}) therefore connect correlation coefficients directly measurable from the EBL--galaxy cross-correlation to parameters of our BCLF model, which can be individually constrained by solving these equations. Although we will focus on the analysis of the BCLF hereafter, for completeness, in Appendix~\ref{appendix:taylor} we also derive the mean and variance of the luminosity ratio, $L_\mathrm{H\alpha}/L_{U}$, as two examples of other potentially useful measures of the BCLF and thus the star formation burstiness. 

\subsection{Relationship between the real-space zero-lag cross-correlation and the Fourier-space cross-bispectrum}  \label{sec:realvsfourier}

Here, before presenting the full cross-correlation analysis framework in Fourier space, we start with a demonstration of how the Poisson-noise cross-bispectrum to be analyzed relates to the zero-lag cross-correlation (i.e., stacking) in real space, which might be more intuitive to understand as a well-established method to probe astrophysics beyond the reach of individually targeted observations \cite[see e.g.,][for a recent stacking analysis of the dust-obscured star formation in high-$z$ galaxies]{Viero_2022}. By showing that they are essentially equivalent, we aim to build up the physical intuition to comprehend details of the full, Fourier-space treatment to be described in Section~\ref{sec:fourierspace}. 

To demonstrate the equivalence of cross-correlation analyses performed in Fourier and real spaces, it is sufficient to compare the signal-to-noise ratios (S/N) derived in both cases as a measure of the information available. For a zero-lag cross-correlation of intensity maps $j$ and $k$ with galaxies in real space, in the Poisson-noise dominated limit, the S/N scales as
\begin{equation}
\left( \mathrm{\frac{S}{N}} \right)_\mathrm{rs} \sim \left(N_\mathrm{gal}\right)^{1/2} \left(\frac{\langle \nu I^j_\nu \rangle}{\sigma^{j}_\mathrm{pix,N}}\right) \left(\frac{\langle \nu I^k_\nu \rangle}{\sigma^{k}_\mathrm{pix,N}}\right) \frac{\langle L_j L_k \rangle}{\langle L_j \rangle \langle L_k \rangle},
\label{eq:snr_real}
\end{equation}
which is a product of the cross-correlation coefficient, the S/N per pixel of the intensity maps, and a scaling factor for the noise reduction when ``stacking'' on $N_\mathrm{gal}$ galaxies. Using definitions of cross-bispectrum and its uncertainty to be introduced in Section~\ref{sec:fourierspace}, we can show that the S/N of cross-bispectrum $B^{ijk}_{\ell}$ defined in Fourier space \textit{resembles equation~(\ref{eq:snr_real}) in the Poisson-noise limit}. Specifically, we have (see Section~\ref{sec:model:unc} for details)
\begin{align}
\left( \mathrm{\frac{S}{N}} \right)^2_{\times} & = \left( \frac{B^{ijk}_{\ell,\mathrm{P}}}{\delta B^{ijk}_{\ell,\mathrm{P}}} \right)^2 \approx \sum_{\ell_1, \ell_2, \ell_3} \frac{\left[B^{ijk}_{\ell}(\ell_1, \ell_2, \ell_3)\right]^2}{C^{i}_{\ell}(\ell_1) C^{j}_{\ell}(\ell_2) C^{k}_{\ell}(\ell_3)} \Omega_\mathrm{s} \ell_\mathrm{max} \Delta \ell_1 \Delta \ell_2 \Delta \ell_3 \nonumber \\
& \approx \ell^4_\mathrm{max} \Omega_\mathrm{s} \frac{\left( B^{ijk}_{\ell,\mathrm{P}} \right)^2}{C^{i}_{\ell,\mathrm{P}} C^{j}_{\ell,\mathrm{P}} C^{k}_{\ell,\mathrm{P}}}, 
\label{eq:snr_fourier}
\end{align}
where the approximation $N_\mathrm{trip} \approx \ell_\mathrm{max} \Omega^2_\mathrm{s} \Delta \ell_1 \Delta \ell_2 \Delta \ell_3 \approx \ell^4_\mathrm{max} \Omega^2_\mathrm{s}$ is applied. Note that here $\ell_\mathrm{max} \approx \theta_\mathrm{pix}^{-1}$, where $\theta_\mathrm{pix}$ is the pixel size in steradian, and $\Omega_\mathrm{s}$ is the survey size. As will be shown in Section~\ref{sec:model:unc}, we can write the angular power spectra as $C^{i}_{\ell,\mathrm{P}} = \Omega_\mathrm{s} N^{-1}_\mathrm{gal}$, $C^{j}_{\ell,\mathrm{P}} = \left(\sigma^{j}_\mathrm{pix,N}\right)^2 \theta^2_\mathrm{pix}$, and $C^{k}_{\ell,\mathrm{P}} = \left(\sigma^{k}_\mathrm{pix,N}\right)^2 \theta^2_\mathrm{pix}$, whereas the cross-bispectrum scales as
\begin{equation}
B^{ijk}_{\ell,\mathrm{P}} \propto \langle \nu I_\nu^{j} \rangle \langle \nu I_\nu^{k} \rangle \frac{\langle L_j L_k \rangle}{\langle L_j \rangle \langle L_k \rangle}.
\end{equation}
Putting together, we can recover the form of equation~(\ref{eq:snr_real}) from equation~(\ref{eq:snr_fourier}). 

Therefore, we stress that, while measuring a zero-lag cross-correlation in real space is mathematically equivalent to measuring a Poisson-noise cross-bispectrum in Fourier space, we choose to work in Fourier space below given practical considerations in observational data analysis that favor it as a more robust and unbiased method. For example, the finite angular and spectral resolution of SPHEREx imply that the pure zero-lag cross-correlation is not strictly observable. The separation between the clustering contributions and Poisson fluctuations is then more transparent in Fourier space, as are the treatment of the beam, spectral resolution, foreground contamination and pixel noise, while the analysis may also be more easily generalized to incorporate clustering terms. 

\subsection{The EBL--galaxy cross-correlation: signals and errors} \label{sec:fourierspace}

\subsubsection{Cross-power spectra and cross-bispectra} \label{sec:psbs}

Following \citet{CC_2022}, we can write the cross-power spectra between H$\alpha$/UV emission and galaxies in the Poisson-noise limit as
\begin{equation}
C^\mathrm{H\alpha,g}_\mathrm{\ell,P} = \frac{1}{\sigma_\mathrm{g}} \Delta z_\mathrm{g} \frac{d \nu I_{\nu}}{d z} \Big|_\mathrm{H\alpha}
\end{equation}
and
\begin{equation}
C^{U,\mathrm{g}}_\mathrm{\ell,P} = \frac{1}{\sigma_\mathrm{g}} \Delta z_\mathrm{g} \frac{d \nu I_{\nu}}{d z} \Big|_{U}.
\end{equation}
By analogy to the definition of cross-power spectra, three fields (two factors of intensity map and one factor of galaxy distribution) are required to calculate ensemble averages involving the second moment of luminosity, $\langle \mathcal{O}(L^2) \rangle$. We therefore define the cross-bispectrum as an integral of the differential flux densities $d (\nu I_{\nu}) / d z$ of H$\alpha$ and UV emission (which themselves are mass integrals over the galaxy population described by the stellar mass function $\psi(M_{*})$) over redshift, conditioned on the subgroup of galaxies selected by stellar mass. When a narrow redshift range $\Delta z_\mathrm{g} \ll 1$ is considered, the redshift integral $\int_{\Delta z_\mathrm{g}} F(z) d z$ can be approximated as $F(z_\mathrm{g}) \Delta z_\mathrm{g}$, which simplifies the calculations. 

For H$\alpha$ (line) and UV (continuum) emission, we can write their differential flux densities as\footnote{Note that we omit the convolution with the conditional PDFs of the luminosities in the two expressions below for brevity, but include them in the full expressions for $B_{\ell,\mathrm{P}}$ below.}
\begin{align}
\frac{d \nu I_{\nu}}{d z} \Big|_\mathrm{H\alpha} & = \frac{1}{\Delta z_\mathrm{H\alpha}} \int_{M_\mathrm{*,min}}^{M_\mathrm{*,max}} d M_* \psi(M_*) \frac{\nu L_\mathrm{H\alpha}}{4\pi D^2_L} \frac{d\chi}{d\nu} D^2_{A,\mathrm{com}} \nonumber \\
& \simeq \frac{1}{\Delta z_\mathrm{g}} \frac{c/H(z_\mathrm{g})}{4\pi (1+z_\mathrm{g})} \int d M_* \psi(M_*) L_\mathrm{H\alpha}~,
\label{eq:fd_ha}
\end{align}
\begin{align}
\frac{d \nu I_{\nu}}{d z} \Big|_{U} & = \int_{M_\mathrm{*,min}}^{M_\mathrm{*,max}} d M_* \psi(M_*) \frac{L_{U}}{4\pi D^2_L} \frac{d \chi}{d z} D^2_{A,\mathrm{com}} \nonumber \\
& = \frac{c/H(z_\mathrm{g})}{4\pi (1+z_\mathrm{g})^2} \int d M_* \psi(M_*) L_{U}~,
\label{eq:fd_uv}
\end{align}
and the density of galaxies (per unit solid angle) is
\begin{equation}
\sigma_\mathrm{g} \simeq \Delta z_\mathrm{g} \frac{d N_\mathrm{g}}{d z d \Omega} = \frac{\chi^2(z_\mathrm{g}) c \Delta z_\mathrm{g}}{H(z_\mathrm{g})} \int d M_* \psi(M_*)~,
\end{equation}
where $H(z)$, $\chi$, $D_{L}$, and $D_{A,\mathrm{com}}=\chi$ are the Hubble parameter, the comoving radial distance, the luminosity distance, and the comoving angular diameter distance, respectively. The $\chi$ gradients are given by $d \chi / d \nu = c(1+z)/[\nu H(z)]$ for the observed frequency $\nu$, and $d \chi / d z = c/H(z)$. We assume $\Delta z_\mathrm{g} \approx \Delta z_\mathrm{H\alpha}=(1+z)/R$ with $R$ being the spectral resolving power. Note that both $L_{U}$ and $L_\mathrm{H\alpha}$ are defined to be non-specific luminosities in units of $\mathrm{erg\,s^{-1}}$ that, to the first order, scale with the SFR and thus $M_{*}$. Unless otherwise specified when the mass integral spans the full range of stellar mass from $M_\mathrm{*,min}=10^{7.5}\,M_{\odot}$ to $M_\mathrm{*,max}=10^{11.5}\,M_{\odot}$ (as in equations~(\ref{eq:fd_ha}) and (\ref{eq:fd_uv})), the stellar mass integral is by default over $\Delta M_*$, which selects the subgroup of galaxies in the stellar mass bin of interest. 

With equations~(\ref{eq:fd_ha}) and (\ref{eq:fd_uv}), the Poisson-noise-limit cross-bispectrum of the H$\alpha$ line, $U$-band continuum, and galaxy fields can be written as
\begin{align}
B^{\times}_{\ell, \mathrm{P}} \equiv &\ B^{\mathrm{H\alpha}, U, \mathrm{g}}_{\ell, \mathrm{P}} \nonumber \\
%= &\ \frac{d (\nu I_{\nu})/d z |_\mathrm{H\alpha}}{\int d M_* \psi(M_*) L_\mathrm{H\alpha}} \frac{d (\nu I_{\nu})/d z |_\mathrm{UV}}{\int d M_* \psi(M_*) L_{U}} \frac{1}{\int \psi(M_*) d M_*} \nonumber \\
%& \times \frac{H^2(z_\mathrm{g})}{c^2\chi^4(z_\mathrm{g})} \int d M_* \psi(M_*) L_\mathrm{H\alpha} L_{U} \nonumber \\
= &\ \frac{1}{\sigma_\mathrm{g}} \int d z \frac{H(z)}{c \chi^2(z) \Delta z_\mathrm{H\alpha|g}} \int d M_{*} \psi(M_{*}) \times \nonumber \\ 
& \left[ \frac{\nu L_\mathrm{H\alpha}}{4\pi D^2_L} \frac{d\chi}{d\nu} D^2_{A,\mathrm{com}} \right] \left[ \frac{L_{U}}{4\pi D^2_L} \frac{d \chi}{d z} D^2_{A,\mathrm{com}} \right] \Phi(L_\mathrm{H\alpha},L_{U}|M_*) \nonumber \\
= &\ \frac{c \int d M_* \Phi(L_\mathrm{H\alpha},L_{U}|M_*) \psi(M_{*}) L_\mathrm{H\alpha} L_{U}}{16\pi^2 \sigma_\mathrm{g} H(z_\mathrm{g}) (1+z_\mathrm{g})^3 \chi^2(z_\mathrm{g})}~,
\label{eq:b_UVhag}
\end{align}
where $\Delta z_\mathrm{H\alpha|g} \approx \Delta z_\mathrm{H\alpha} \approx \Delta z_\mathrm{g}$ denotes the redshift range over which galaxy and emission intensity fields overlap. Similarly, for the $\langle L^2_{U} \rangle$ and $\langle L^2_\mathrm{H\alpha} \rangle$ (auto-correlation) terms in the denominator of equation~(\ref{eq:r}), we have
\begin{align}
B^{U, U, \mathrm{g}}_{\ell, \mathrm{P}} = &\ \frac{1}{\sigma_\mathrm{g}} \int d z \frac{H(z)}{c \chi^2(z)} \int d M_* \psi(M_*,z) \nonumber \\ 
& \times \left[ \frac{L_{U}}{4\pi D^2_L} \frac{d \chi}{d z} D^2_{A,\mathrm{com}} \right]^2 \Phi(L_{U}|M_*) \nonumber \\
= &\ \frac{c \Delta z_\mathrm{g}}{H(z_\mathrm{g})} \frac{\int d M_* \Phi(L_{U}|M_*) \psi(M_*,z_\mathrm{g}) L^2_{U}}{16\pi^2 \sigma_\mathrm{g} (1+z_\mathrm{g})^4 \chi^2(z_\mathrm{g})}~,
\label{eq:b_UVUVg}
\end{align}
and
\begin{align}
B^\mathrm{H\alpha, H\alpha, g}_{\ell, \mathrm{P}} = &\ \frac{1}{\sigma_\mathrm{g}} \int d z \frac{H(z)}{c \chi^2(z)} \frac{1}{\Delta z^2_\mathrm{H\alpha}} \int d M_* \psi(M_*,z) \nonumber \\ 
& \times \left[ \frac{\nu L_\mathrm{H\alpha}}{4\pi D^2_L} \frac{d \chi}{d \nu} D^2_{A,\mathrm{com}} \right]^2 \Phi(L_\mathrm{H\alpha},L_{U}|M_*) \nonumber \\
= &\ \frac{1}{\sigma_\mathrm{g}} \frac{c}{H(z_\mathrm{g}) \Delta z_\mathrm{g}} \frac{\int d M_* \psi(M_*,z_\mathrm{g}) L^2_\mathrm{H\alpha}}{16\pi^2 (1+z_\mathrm{g})^2 \chi^2(z_\mathrm{g})}~.
\label{eq:b_HaHag}
\end{align}

\begin{figure}
    \centering
	\includegraphics[width=\columnwidth]{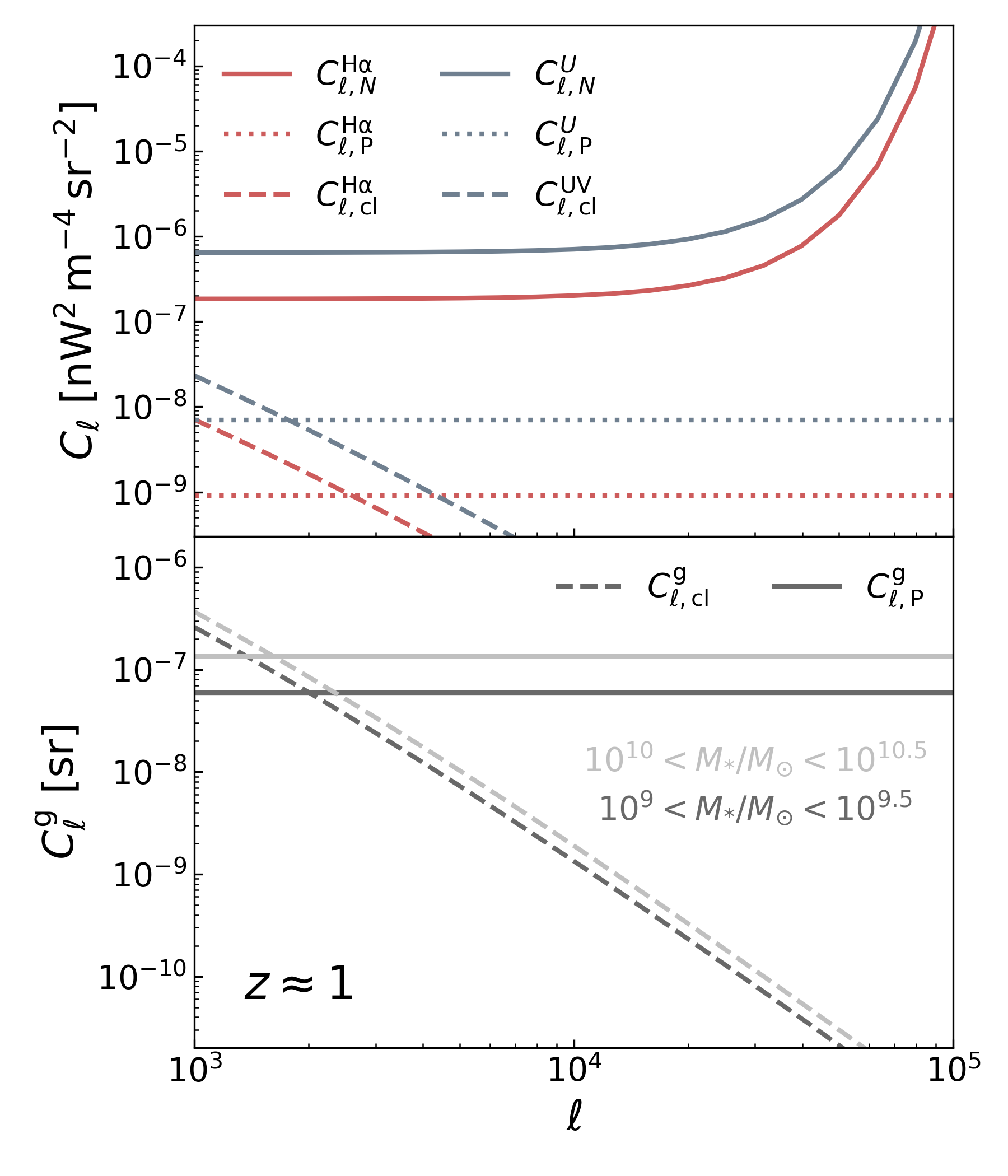}
    \caption{A comparison of the error budget for $C_{\ell}$ of H$\alpha$ and UV emission (top panel), as well as the galaxy distribution (bottom panel) at $z\approx1.5$. At high multipoles $\ell \geq 10^4$, uncertainties of the intensity (galaxy) power spectra are strongly dominated by the instrument noise $C_{\ell, N}$ (Poisson noise $C^\mathrm{g}_{\ell, \mathrm{P}}$). Note that, unlike in the bottom panel, the sample variances in the top panel are evaluated by integrating over the full range of stellar mass $[M_\mathrm{*,min}, M_\mathrm{*,max}]$.}
    \label{fig:errorbudget}
\end{figure}

\subsubsection{Uncertainties on cross-power spectra and cross-bispectra} \label{sec:model:unc}

For the cross-power spectrum between an intensity map $i$ (H$\alpha$ or $U$-band intensity map here) and galaxies, the uncertainty for a given multipole moment $\ell$ binned in a width of $\Delta \ell$ can be expressed as
\begin{equation}
\left(\delta C^{i, \mathrm{g}}_{\ell, \mathrm{P}}\right)^2 = \frac{1}{f_\mathrm{sky}(2\ell+1)\Delta \ell} \left[ \left(C^{i, \mathrm{g}}_{\ell, \mathrm{P}}\right)^2 + C^{i}_{\ell,\mathrm{N}} C^\mathrm{g}_{\ell,\mathrm{P}} \right],
\label{eq:cps_unc}
\end{equation}
where $f_\mathrm{sky}$ is the sky covering fraction and we assume here that auto-correlations of the intensity map and galaxies are dominated by the instrument noise and the Poisson noise, respectively, on the small scales considered in our analysis. In practice, to obtain the net effective uncertainty of the cross-power spectrum, we further scale down equation~(\ref{eq:cps_unc}) by a factor of 300 to approximate the gain in sensitivity from binning together modes over $10^4 < \ell < 10^5$. This renders S/N of $C^{\mathrm{H\alpha}, \mathrm{g}}_{\ell, \mathrm{P}}$ (or $C^{U, \mathrm{g}}_{\ell, \mathrm{P}}$) substantially higher than that of $B^\mathrm{H\alpha, H\alpha, g}_{\ell, \mathrm{P}}$ (or $B^{U, U, \mathrm{g}}_{\ell, \mathrm{P}}$), as will be detailed below, and therefore the S/N of auto-correlation coefficients $r^\mathrm{g}_\mathrm{H\alpha,P}$ (or $r^\mathrm{g}_{U,\mathrm{P}}$) can be simply approximated as twice of that of $B^\mathrm{H\alpha, H\alpha, g}_{\ell, \mathrm{P}}$ (or $B^{U, U, \mathrm{g}}_{\ell, \mathrm{P}}$).  

Following \citet{Kayo_2013}, we can write the bispectrum variance in the Gaussian approximation as
\begin{equation}
\mathrm{Var}\left[ B^{ijk}_{\ell}(\ell_1, \ell_2, \ell_3) \right] = \frac{\Omega_\mathrm{s} C_\ell^i(\ell_1)C_\ell^j(\ell_2)C_\ell^k(\ell_3)}{N_\mathrm{trip}(\ell_1, \ell_2, \ell_3)},
\label{eq:varB}
\end{equation}
where $\Omega_\mathrm{s}$ is the total survey area over which EBL and galaxy surveys overlap ($\Omega_\mathrm{s} \approx 5.5\,\mathrm{sr}$ for SPHEREx and Rubin/LSST), and the number of triplets that form closed triangles in Fourier space $N_\mathrm{trip}(\ell_1, \ell_2, \ell_3) \equiv \sum_{\boldsymbol{q}_i; q_i \in \ell_i} \Delta \boldsymbol{q}_{123}$, which can be approximated in the limit of large multipole bins as
\begin{equation}
N_\mathrm{trip} \simeq \frac{\Omega_\mathrm{s}^2 \ell_1 \ell_2 \ell_3 \Delta \ell_1 \Delta \ell_2 \Delta \ell_3 / 2\pi^3}{\sqrt{2\ell^2_1 \ell^2_2 + 2\ell^2_1 \ell^2_3 + 2\ell^2_2 \ell^2_3 - \ell_1 - \ell_2 - \ell_3}}.
\end{equation}
Each of the three angular auto-power spectra in the numerator of equation~(\ref{eq:varB}) has contributions from clustering\footnote{For simplicity, we ignore the nonlinear clustering whose impact on scales smaller than $\ell\sim10^4$ is expected to be subdominant to that of the Poisson noise \cite[see e.g.,][]{CB_2022}.}, Poisson noise, and instrument noise (for intensity maps of H$\alpha$ and UV emission) whose relatively importance varies across $\ell$. Specifically, assuming Limber approximation and narrow redshift range $\Delta z_\mathrm{g} \ll 1$, we have \citep{CC_2022}
\begin{equation}
C^\mathrm{g}_{\ell,\mathrm{cl}}(\ell) = \frac{H(z_\mathrm{g}) \langle b \rangle^2_\mathrm{g}(z_\mathrm{g})}{\Delta z_\mathrm{g} c \chi^2(z_\mathrm{g})} P_{\delta\delta}\left[ k=\frac{\ell}{\chi(z_\mathrm{g})}, z_\mathrm{g}\right]
\end{equation}
and
\begin{equation}
C^\mathrm{g}_{\ell,\mathrm{P}} = \left( \frac{d N_\mathrm{g}}{d \Omega} \right)^{-1} = \sigma^{-1}_\mathrm{g}
\end{equation}
for the auto-power spectrum of galaxies, where $\langle b \rangle_\mathrm{g}$ is the galaxy bias averaged over the ensemble of galaxies in the stellar mass bin of width $\Delta M_{*}$ (see Appendix~\ref{appendix:bias} for a more detailed description of the various bias factors involved) and $P_{\delta\delta}$ is the dark matter power spectrum. Similarly, for H$\alpha$ and UV emission, the auto-power spectra are
\begin{align}
C^\mathrm{H\alpha}_{\ell,\mathrm{cl}}(\ell) = & \int d z \frac{H(z)}{c \chi^2(z)} b^2_\mathrm{H\alpha}(z) \left[ \frac{d \nu I_{\nu}}{d z} \Big|_\mathrm{H\alpha}(z) \right]^2 \nonumber \\
& \times P_{\delta\delta}\left[ k=\frac{\ell}{\chi(z)}, z \right],
\end{align}
\begin{align}
C^\mathrm{H\alpha}_{\ell,\mathrm{P}} = & \int d z \frac{H(z)}{c \chi^2(z)} \frac{1}{\Delta z^2_\mathrm{H\alpha}} \int_{M_\mathrm{*,min}}^{M_\mathrm{*,max}} d M_* \psi(M_*,z) \nonumber \\ 
& \times \left[ \frac{\nu L_\mathrm{H\alpha}}{4\pi D^2_L} \frac{d \chi}{d \nu} D^2_{A,\mathrm{com}} \right]^2 \Phi(L_\mathrm{H\alpha},L_{U}|M_*),
\label{eq:acps_ha}
\end{align}
\begin{align}
C^{U}_{\ell,\mathrm{cl}}(\ell) = & \int d z \frac{H(z)}{c \chi^2(z)} b^2_{U}(z) \left[ \frac{d \nu I_{\nu}}{d z} \Big|_{U}(z) \right]^2 \nonumber \\
& \times P_{\delta\delta}\left[ k=\frac{\ell}{\chi(z)}, z \right],
\end{align}
and
\begin{align}
C^{U}_{\ell,\mathrm{P}} = & \int d z \frac{H(z)}{c \chi^2(z)} \int_{M_\mathrm{*,min}}^{M_\mathrm{*,max}} d M_* \psi(M_*,z) \nonumber \\ 
& \times \left[ \frac{L_{U}}{4\pi D^2_L} \frac{d \chi}{d z} D^2_{A,\mathrm{com}} \right]^2 \Phi(L_{U}|M_*).
\label{eq:acps_uv}
\end{align}

\begin{figure*}
    \centering
	\includegraphics[width=\columnwidth]{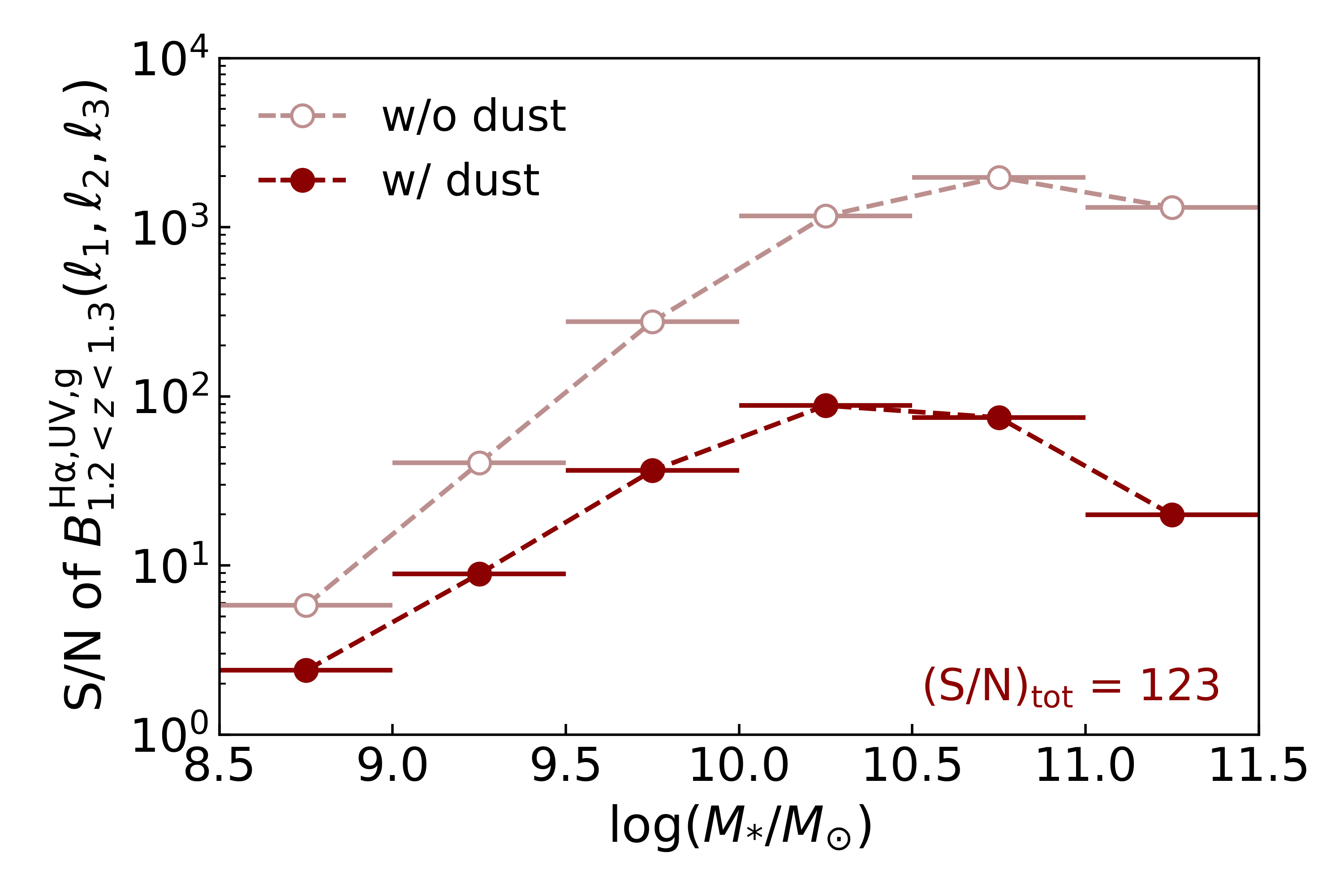}
	\includegraphics[width=\columnwidth]{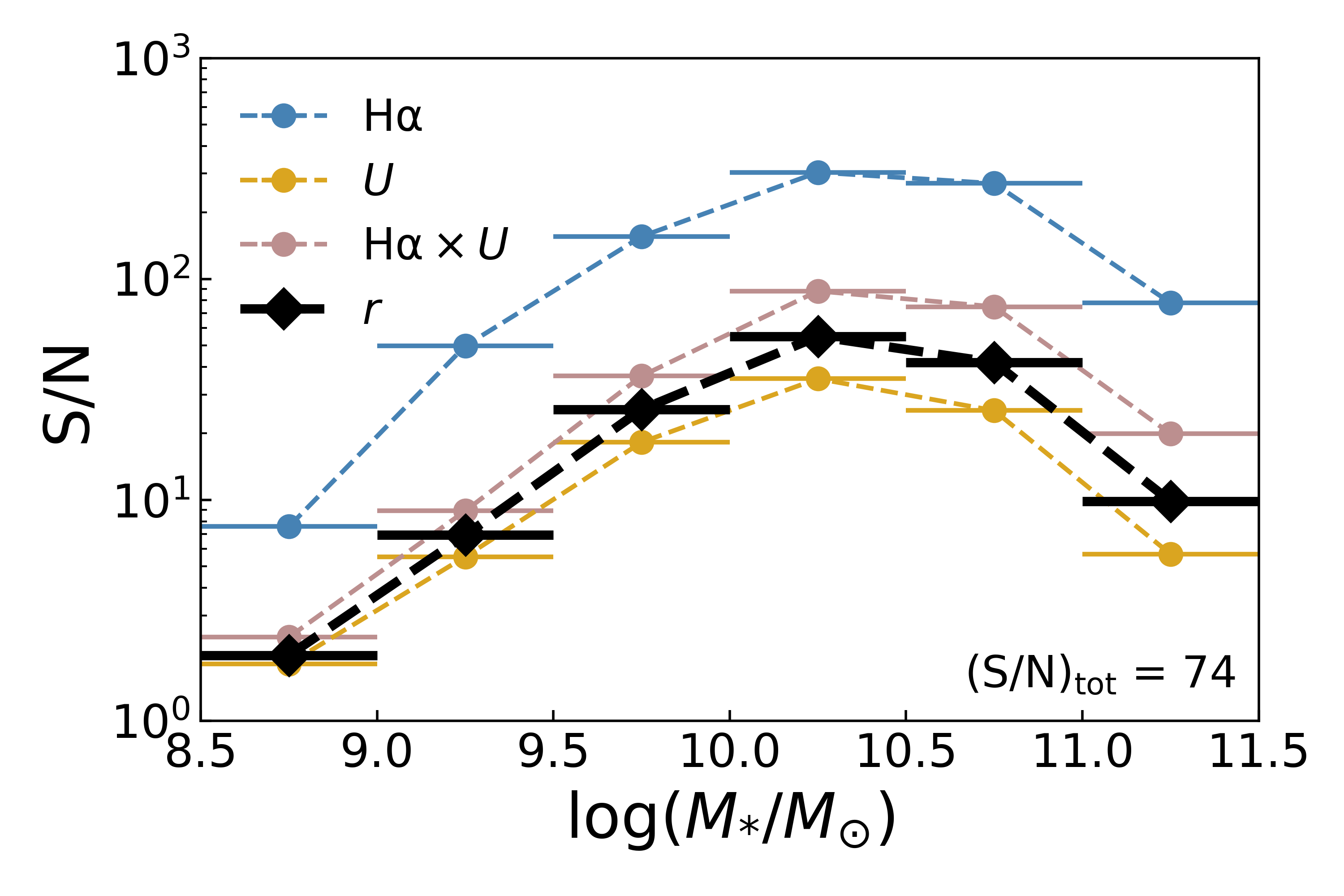}
    \caption{Left: S/N of the Poisson-noise cross-bispectra of H$\alpha$, UV, and galaxies at $z\approx1.25$ in different stellar mass bins, before and after including dust attenuation. The fiducial model (Model~I) is assumed and the total S/N is quoted for the sum over all stellar mass bins. Right: a comparison of the detectability of the cross-correlation coefficient, $r^\mathrm{g}_\mathrm{\times,P}$ (black), as well as its 3 components, namely $B^{\mathrm{H\alpha},U,\mathrm{g}}_\mathrm{\ell,P}$ (red), $B^{\mathrm{H\alpha},\mathrm{H\alpha},\mathrm{g}}_\mathrm{\ell,P}$ (blue), and $B^{U,U,\mathrm{g}}_\mathrm{\ell,P}$ (yellow). The fiducial model (Model~I) is assumed, after including dust attenuation. All the data displayed here are evaluated for a single redshift interval, without redshift binning (see Section~\ref{sec:results:cross-correlations}).}
    \label{fig:b_snr}
\end{figure*}

As shown by Fig.~\ref{fig:errorbudget}, on small scales the Poisson noise and instrument noise dominate the angular power spectra of galaxies and emission fields, respectively. Therefore, we take $C^{i}_{\ell} = C^\mathrm{g}_{\ell,\mathrm{P}}=\sigma^{-1}_\mathrm{g}$, $C^{j}_{\ell} = C^\mathrm{H\alpha}_{\ell,\mathrm{N}}=\sigma_\mathrm{pix,N}^2|_{\lambda_\mathrm{H\alpha}(1+z)}\Omega_\mathrm{pix}e^{\Omega_\mathrm{pix}\ell^2}$ and, $C^{k}_{\ell} = C^{U}_{\ell,\mathrm{N}}=\sigma_\mathrm{pix,N}^2|_{\lambda_{U}(1+z)}\Omega_\mathrm{pix}e^{\Omega_\mathrm{pix}\ell^2}$, where $\sigma_\mathrm{pix,N}$ is the projected surface brightness sensitivity of the SPHEREx all-sky survey\footnote{See the public data product of surface brightness sensitivity available at \url{https://github.com/SPHEREx/Public-products/blob/master/Surface_Brightness_v28_base_cbe.txt}}. 
To estimate the detectability of the bispectrum in terms of its total signal-to-noise ratio (S/N), we adopt a universal bin size of $\Delta \ell = 1000$ and sum the S/N of individual $\ell$ bins over $\ell_\mathrm{min} = 10^4$ to $\ell_\mathrm{max} = 10^5$ where the angular power spectra are well within the Poisson-noise-dominated regime, namely
\begin{equation}
\left( \frac{\mathrm{S}}{\mathrm{N}} \right)^2_{\times} = \sum^{\ell_\mathrm{max}+\frac{\Delta \ell}{2}}_{\{\ell_1,\ell_2,\ell_3\}=\ell_\mathrm{min}-\frac{\Delta \ell}{2}} \frac{\left(B^{ijk}_{\ell, \mathrm{P}}\right)^2}{\mathrm{Var}\left[B^{ijk}_{\ell}(\ell_1, \ell_2, \ell_3)\right]}.
\end{equation}
Finally, from the definition of $r^\mathrm{g}_\mathrm{\times,P}$, we have
\begin{equation}
\left( \mathrm{\frac{S}{N}} \right)^{-2}_{r_{\times}} = \left( \mathrm{\frac{S}{N}} \right)^{-2}_{\times} + \frac{1}{4} \left[ \left( \mathrm{\frac{S}{N}} \right)^{-2}_\mathrm{H\alpha} + \left( \mathrm{\frac{S}{N}} \right)^{-2}_{U} \right].
\label{eq:snr_error_propagation}
\end{equation}

\section{Results} \label{sec:results}

In this section, we first present the detectability of the various cross-bispectra related to our case study, where we cross-correlate EBL maps of rest-frame H$\alpha$ and UV ($U$-band) emission and photometric galaxies to be observed with SPHEREx and Rubin/LSST, respectively (Section~\ref{sec:results:cross-correlations}). Then, we show the constraints on BCLF model parameters derived from the predicted sensitivity to the correlation coefficients, $r^\mathrm{g}_{\times,\mathrm{P}}$, $r^\mathrm{g}_{\mathrm{H\alpha},\mathrm{P}}$, and $r^\mathrm{g}_{U,\mathrm{P}}$ (Section~\ref{sec:results:model_constraints}). The toy models considered here suffice to forecast the potential for EBL--galaxy cross-correlations to distinguish these limiting cases and thereby shed light on bursty star formation.    

\subsection{Detectability of cross-correlation signals} \label{sec:results:cross-correlations}

In the left panel of Fig.~\ref{fig:b_snr}, we show the predicted detectability of the cross-bispectrum, $B^{\mathrm{H\alpha},U,\mathrm{g}}_{\ell,\mathrm{P}}$, of H$\alpha$ and $U$-band intensity maps measured by the all-sky survey with SPHEREx and photo-$z$ galaxies surveyed by Rubin/LSST in each of the 6 stellar mass bins. The S/N numbers quoted here are evaluated for \textit{a single pair} of spectral channels corresponding to a narrow redshift range of $\Delta z = (1+z)/R$ around $z=1.5$, where $R=41$ is the spectral resolving power of SPHEREx in bands relevant to this study. We note that at the redshifts of interest for this study ($z\sim1.5$--2.5), the adopted $\Delta z$ happens to be comparable to the level of photometric redshift uncertainty expected for the nominal 10-year Rubin/LSST survey, which may be further improved over the course of the survey by the addition of near-IR and UV photometry from other existing/concurrent surveys, such as Roman, Euclid, and SPHEREx \citep{Graham_2018,Graham_2020}. Due to the trade-off between the brightness of sources and the number density of galaxies contributing to the intensity fields and available for cross-correlation, the expected S/N of $B^{\mathrm{H\alpha},U,\mathrm{g}}_{\ell,\mathrm{P}}$ peaks at intermediate mass scales $M_{*} \sim 10^{10.5}\,M_{\odot}$, although a high-significance detection can be achieved in all but the lowest mass bins. Meanwhile, from the comparison between cases with and without dust attenuation, it is clear that the expected detectability of the EBL--galaxy cross-correlation is highly sensitive to the treatment of dust attenuation (especially for massive galaxies that are more dust-rich), which has sometimes been neglected for simplicity in previous work, although dust attenuation will likely reduce the SNR of EBL observations with SPHEREx \cite[e.g.,][]{Gong_2017}. 

In the right panel of Fig.~\ref{fig:b_snr}, we show how the S/N of each bispectrum involved in the definition of $r^\mathrm{g}_{\times,\mathrm{P}}$ can be propagated to obtain the S/N of $r^\mathrm{g}_{\times,\mathrm{P}}$ (see equations~(\ref{eq:r}) and (\ref{eq:snr_error_propagation})). As shown by the comparison, the detectability of $r^\mathrm{g}_{\times,\mathrm{P}}$, from which constraints on $\sigma_{\alpha U}$ (and other BCLF model parameters) are drawn, evolves across the mass bins in a similar way to the bispectra and is mainly set by how well $B^{\mathrm{H\alpha},U,\mathrm{g}}_{\ell,\mathrm{P}}$ can be measured. 

\begin{figure}
    \centering
	\includegraphics[width=\columnwidth]{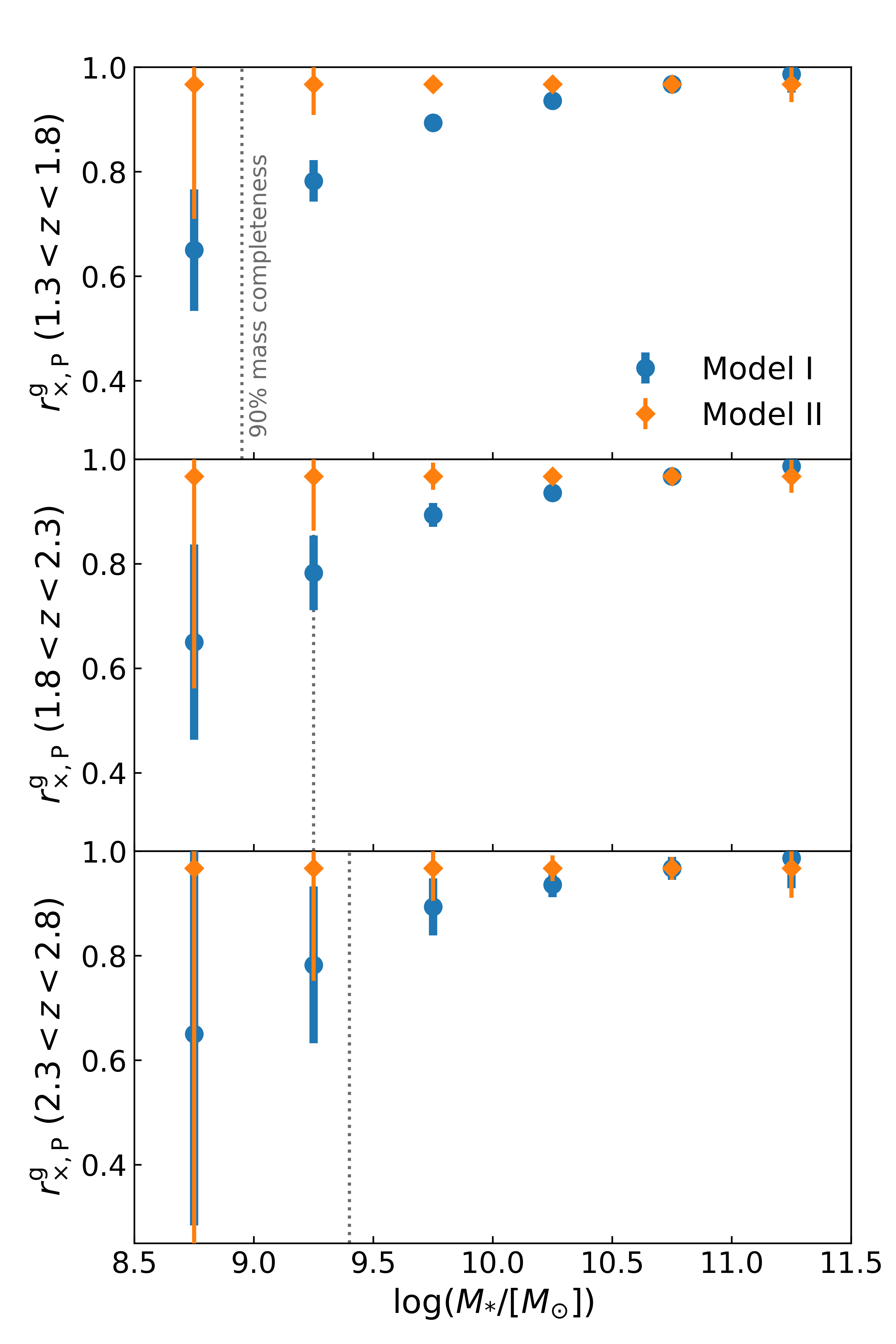}
    \caption{Capability of distinguishing Model~I (blue) and Model~II (orange) implied by the constraints on $r^\mathrm{g}_\mathrm{\times,P}$ in individual stellar mass bins, after binning in redshift. From the top to the bottom, the three panels show the expected constraints evaluated in the three broad redshift bins respectively. The vertical dotted lines indicate the 90\% mass completeness limits of the Rubin/LSST photometric galaxy redshift survey expected at these redshifts (Section~\ref{sec:model_overview}).}
    \label{fig:rconstraints}
\end{figure}

Different from the predicted constraints on the bispectra presented in Fig.~\ref{fig:b_snr}, which are evaluated for a single redshift interval using one pair of spectral channels of SPHEREx, we consider broader redshift bins for measuring the BCLF of $L_\mathrm{H\alpha}$ and $L_{U}$ from the correlation coefficients to optimize the parameter constraints. Specifically, we define 3 redshift bins with bin centers $z_{c}=1.5$, $z_{c}=2.0$, and $z_{c}=2.5$, and bin edges $[1.25, 1.75]$, $[1.75, 2.25]$, and $[2.25, 2.75]$, respectively. We further divide each redshift bin into $\mathcal{N} = 0.5R/(1+z_c)$ redshift intervals with $R=41$, which yields $\mathcal{N} = 8, 7, 6$, respectively. The uncertainties in the correlation coefficients evaluated for $z_c$ and $\Delta z = (1+z_c)/R$ are consequently scaled by a factor of $1/\sqrt{\mathcal{N}}$ to approximate the effect of binning together $\mathcal{N}$ redshift intervals. 

Fig.~\ref{fig:rconstraints} shows the constraints on the cross-correlation coefficient, $r^\mathrm{g}_{\times,\mathrm{P}}$, in each stellar mass bin predicted by Models~I and II in three broad redshift bins as labeled on the vertical axis. With the help of the additional statistical power from redshift binning, we expect cross-correlating EBL maps from SPHEREx with photo-$z$ galaxies from Rubin/LSST to distinguish Model~II from Model~I by detecting the decrease of $r^\mathrm{g}_{\times,\mathrm{P}}$ towards lower stellar masses at high significance up to $z\sim3$. It is noteworthy that even though the difference between the two toy models is modest in intermediate-mass bins, strong evidence for decorrelation may still be obtained thanks to the expected high sensitivity to the bispectra at these mass scales. Detecting such a decorrelation between H$\alpha$ and UV luminosities in low-mass galaxies and characterizing the mass dependence via the EBL--galaxy cross-correlation described can be a smoking gun for an elevated level of bursty star formation, although alternative explanations may exist (see discussion in Section~\ref{sec:discussion}). We note that, for simplicity, instead of estimating the actual galaxy counts taking into account of the mass incompleteness, we show the 90\% mass completeness limit in Fig.~\ref{fig:rconstraints} and note that the constraining power in lower mass bins should therefore be taken as an upper limit due to incompleteness. 

\begin{figure}
    \centering
	\includegraphics[width=0.23\textwidth]{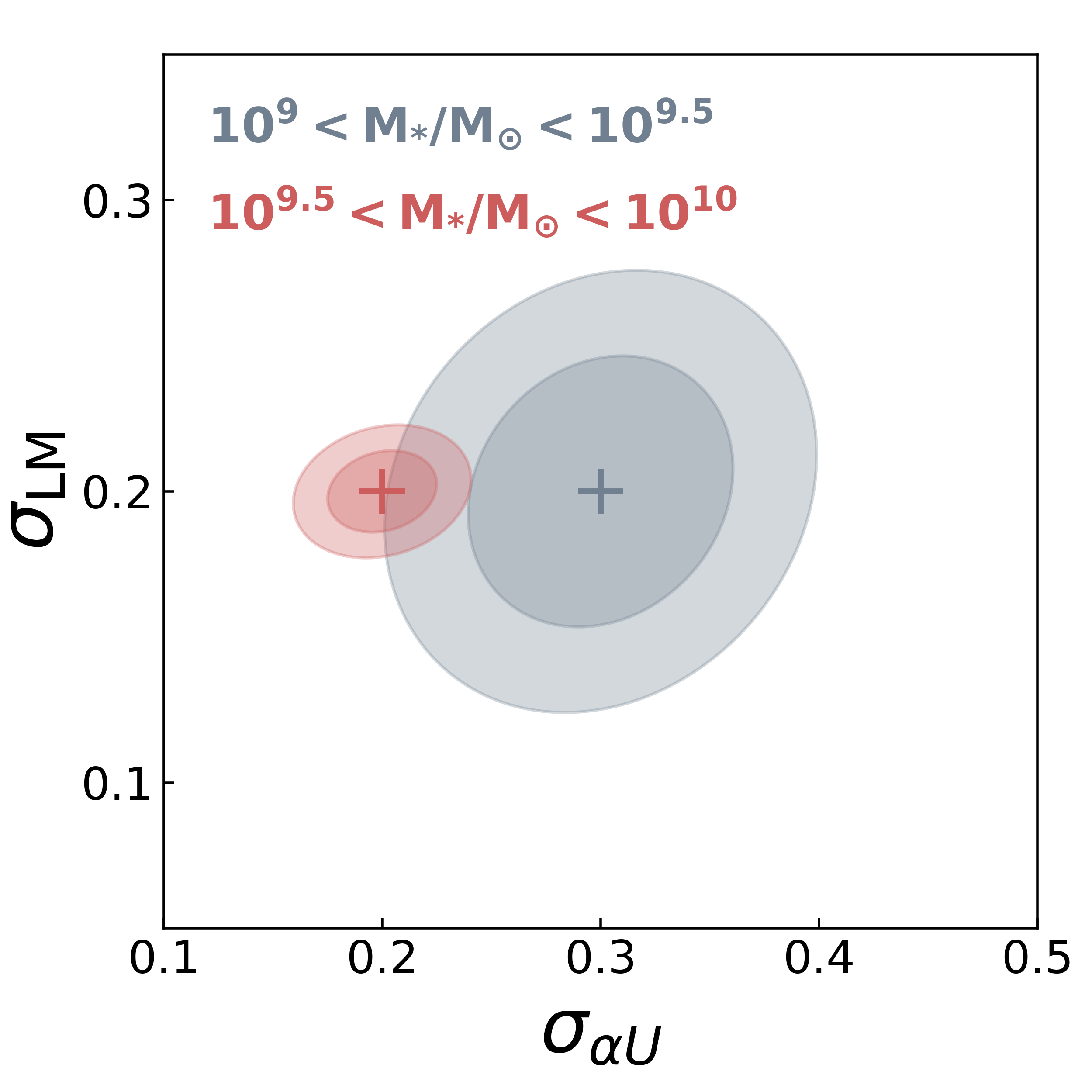}
 	\includegraphics[width=0.23\textwidth]{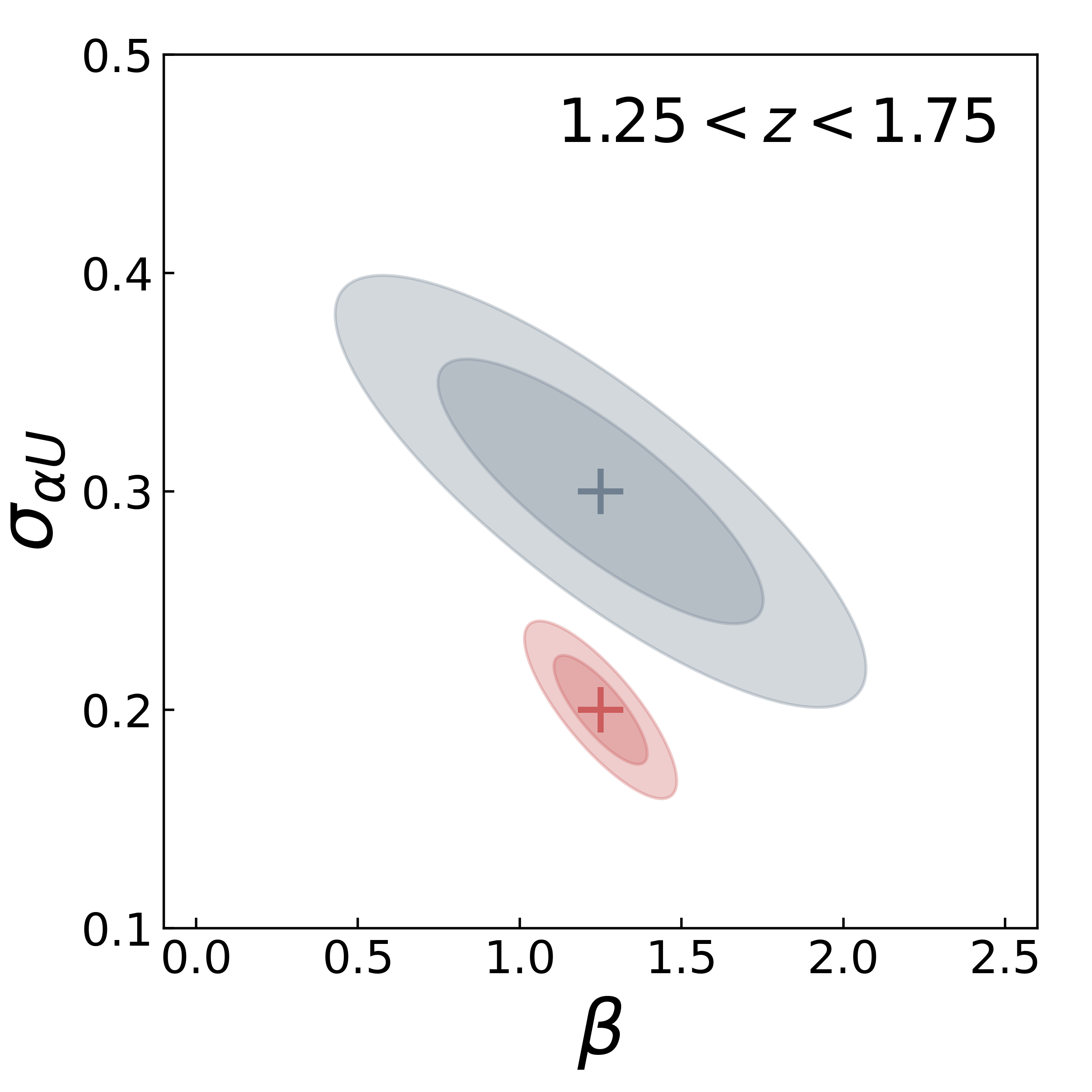}
	\includegraphics[width=0.23\textwidth]{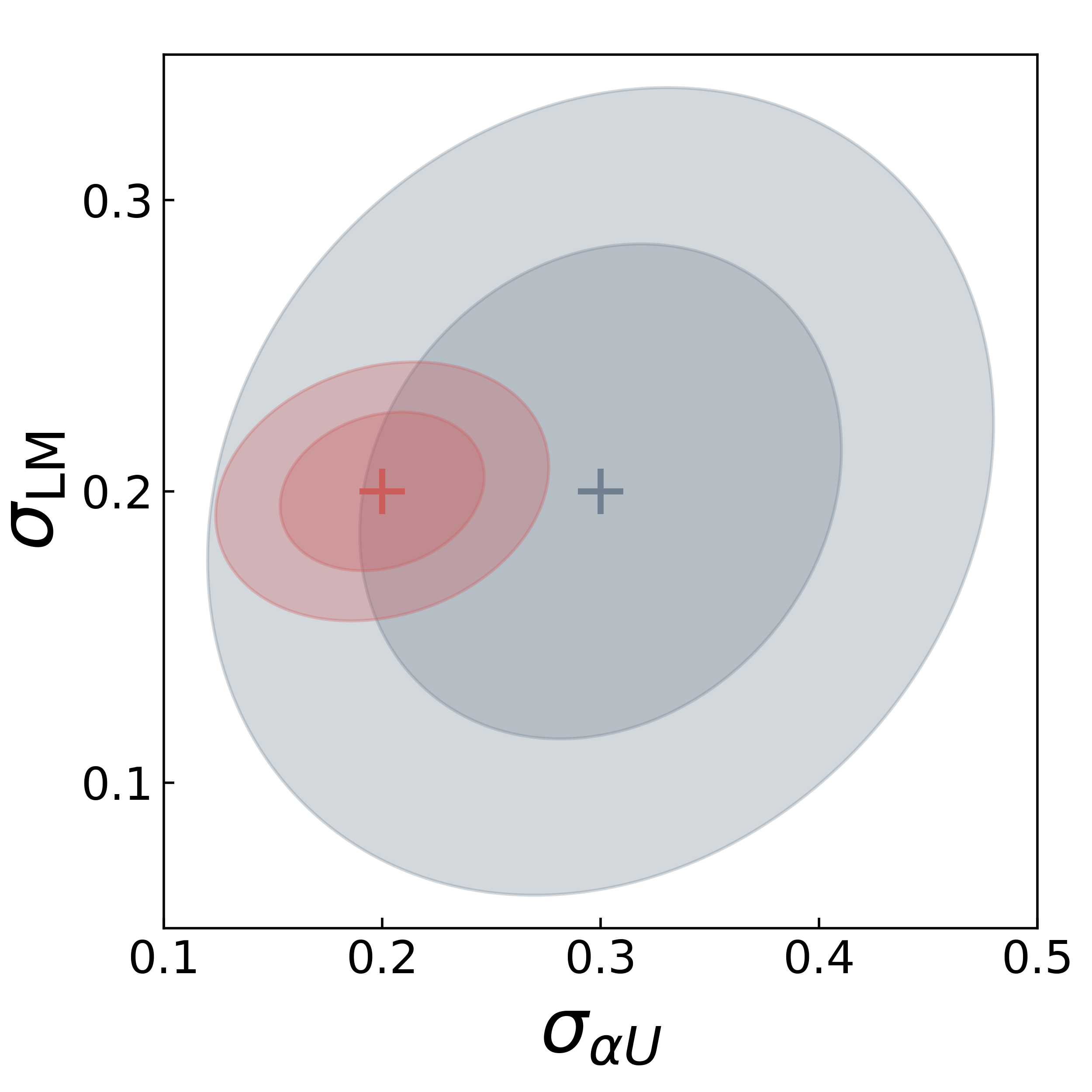}
 	\includegraphics[width=0.23\textwidth]{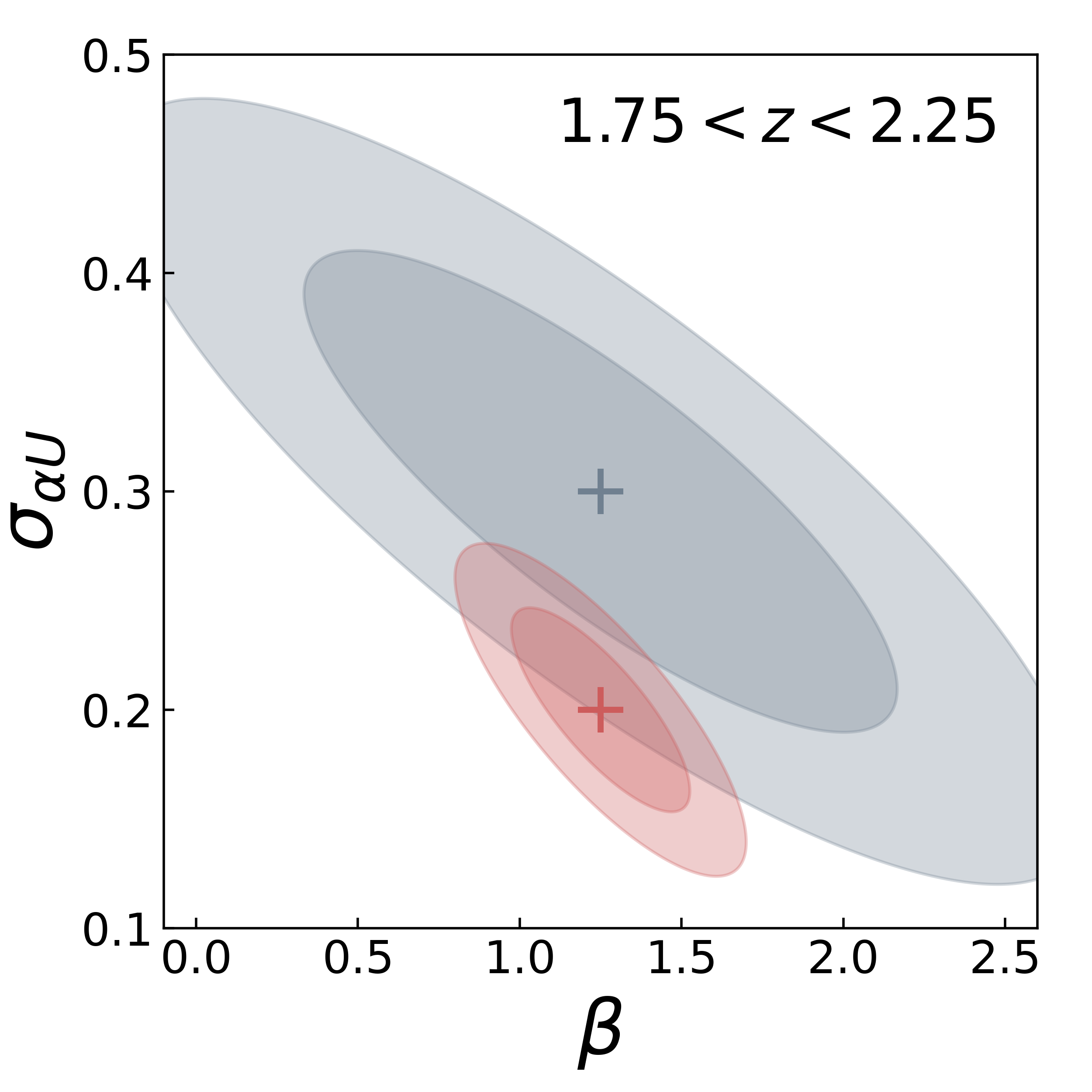}
	\includegraphics[width=0.23\textwidth]{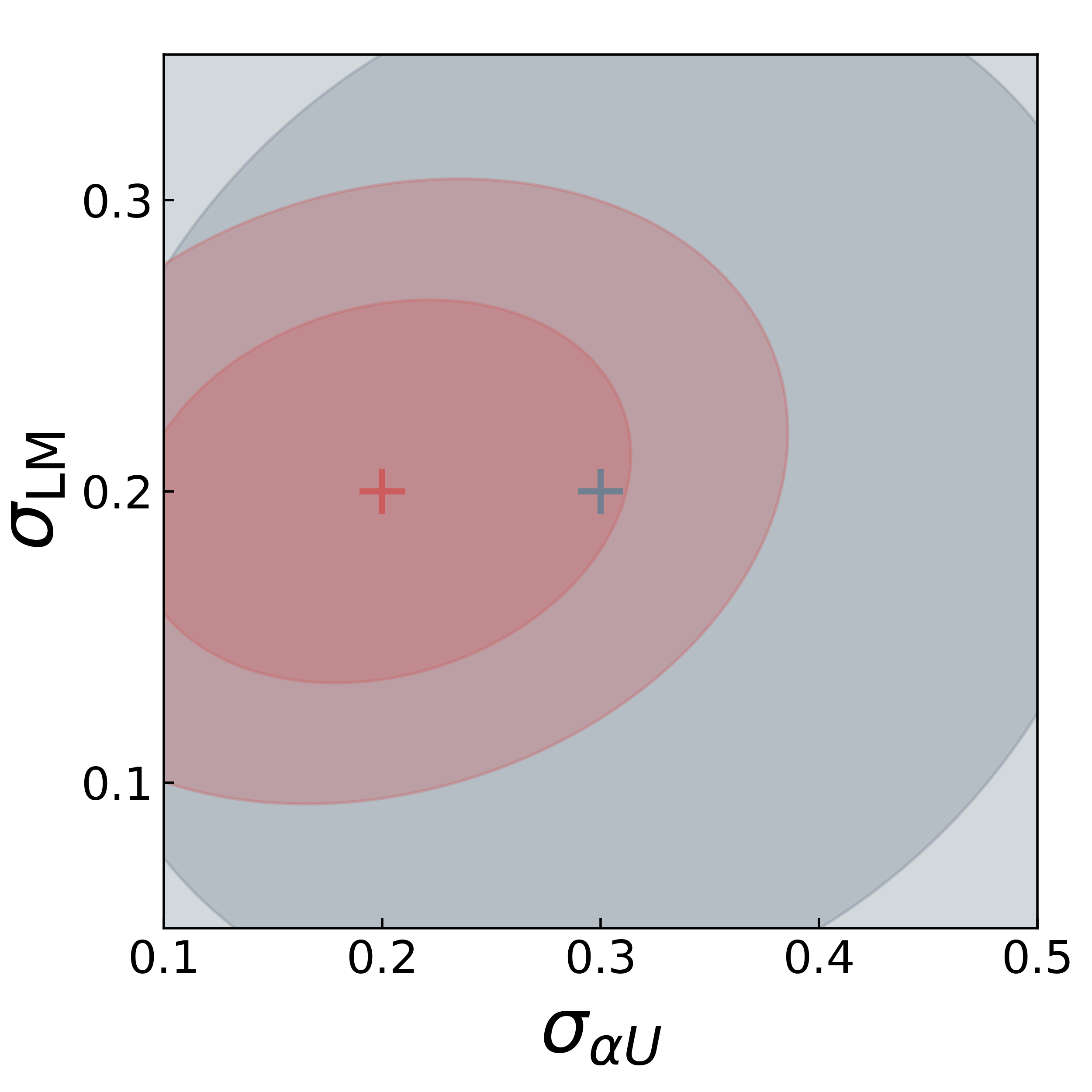}
 	\includegraphics[width=0.23\textwidth]{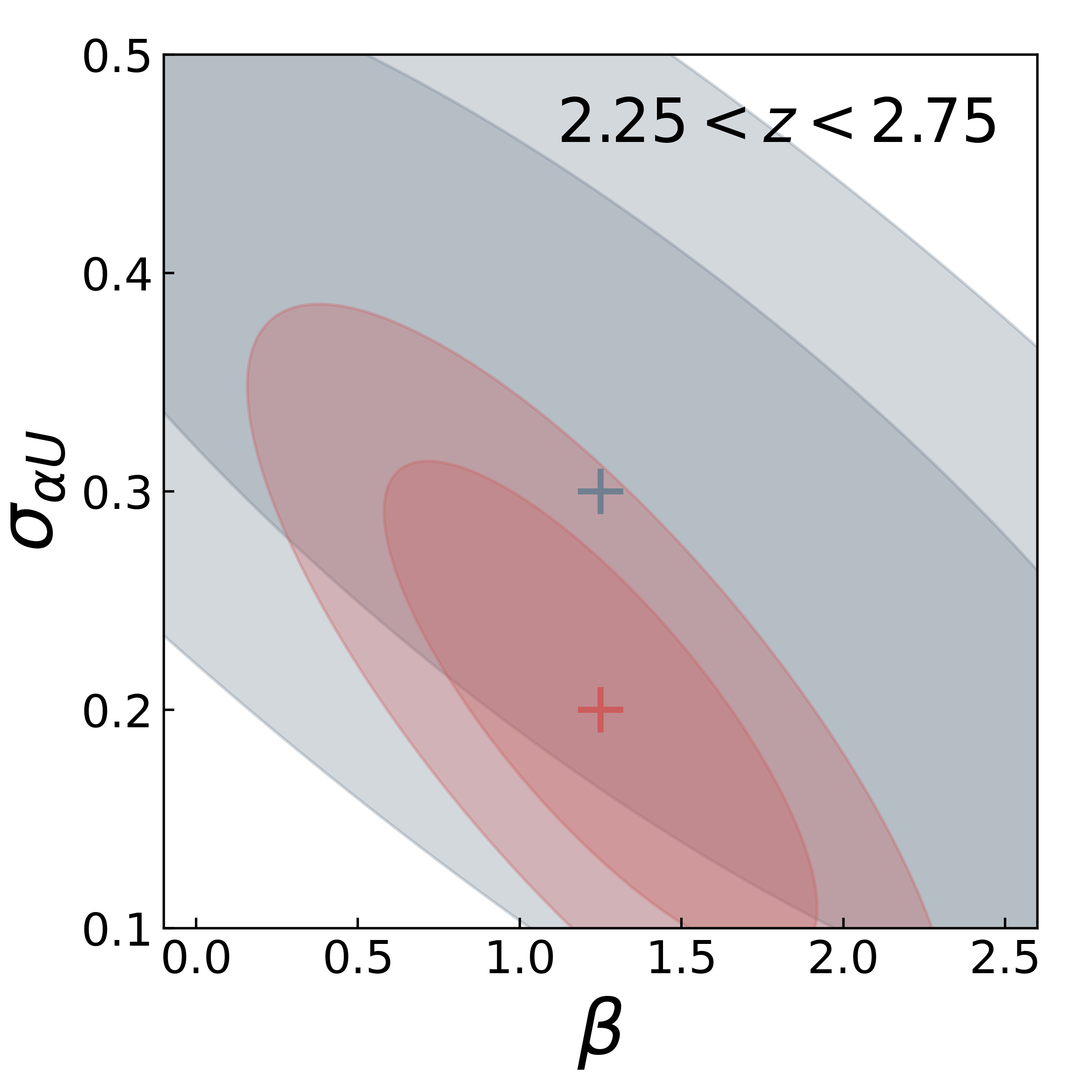}
    \caption{Constraints on the BCLF model parameters $\sigma_{\alpha U}$, $\sigma_\mathrm{LM}$, and $\beta$ in Model~I drawn from the cross-correlation analysis for two example stellar mass bins in different colours and in the three redshift bins from top to bottom. The dark and light shaded ellipses represent the 1-$\sigma$ and 2-$\sigma$ confidence intervals, respectively.}
    \label{fig:param_constraints}
\end{figure}

\subsection{Constraints on BCLF model parameters} \label{sec:results:model_constraints}

From the expected constraints on $r^\mathrm{g}_{\times,\mathrm{P}}$ shown in Fig.~\ref{fig:rconstraints}, together with the similarly derived constraints on the auto-correlation coefficients $r^\mathrm{g}_\mathrm{H\alpha,P}$ and $r^\mathrm{g}_{U,\mathrm{P}}$ (see equations~(\ref{eq:lnrha}) and (\ref{eq:lnruv})), we can directly constrain the H$\alpha$--UV BCLF model assumed. To estimate the parameter constraints, we employ a Fisher matrix formalism, which performs a quadratic expansion around the log-likelihood of the data vector $\hat{\boldsymbol{r}}$, namely
\begin{equation}
F_{ij} = \sum_{k} \frac{1}{\mathrm{var}(r_k)} \frac{\partial r_k}{\partial \theta_i} \frac{\partial r_k}{\partial \theta_2}~,
\end{equation}
with $\boldsymbol{r}(\boldsymbol{\theta}) = \left(r^\mathrm{g}_{\times,\mathrm{P}}(\boldsymbol{\theta}), r^\mathrm{g}_{\mathrm{H\alpha},\mathrm{P}}(\boldsymbol{\theta}), r^\mathrm{g}_{U,\mathrm{P}}(\boldsymbol{\theta})\right)$ being the model vector for $\boldsymbol{\theta} = (\sigma_{\alpha U}, \sigma_\mathrm{LM}, \beta)$. We neglect the covariance between the correlation coefficients, which is likely a reasonable approximation in the instrument-noise-dominated regime relevant to this work (Section~\ref{sec:model:unc}), and no priors are assumed on the parameters. 

\begin{table}
\centering
\caption{Fractional uncertainties in the BCLF model parameters in different redshift and mass bins derived from the diagonal of the inverse of the Fisher matrix.}
\label{tb:fisher_params}
\begin{tabular}{ccccc}
\toprule
Redshift & $\log(M_{*}/M_{\odot})$ & $\sigma_{\alpha U}/\sigma^\mathrm{fid}_{\alpha U}$ & $\beta/\beta^\mathrm{fid}$ & $\sigma_\mathrm{LM}/\sigma^\mathrm{fid}_\mathrm{LM}$ \\
\hline
$1.25$\,<\,$z$\,<\,$1.75$ & 9 -- 9.5 & 0.133 & 0.265 & 0.153 \\
$1.25$\,<\,$z$\,<\,$1.75$ & 9.5 -- 10 & 0.082 & 0.076 & 0.046 \\
$1.75$\,<\,$z$\,<\,$2.25$ & 9 -- 9.5 & 0.242 & 0.482 & 0.280 \\
$1.75$\,<\,$z$\,<\,$2.25$ & 9.5 -- 10 & 0.154 & 0.145 & 0.090 \\
$2.25$\,<\,$z$\,<\,$2.75$ & 9 -- 9.5 & 0.532 & 1.046 & 0.595 \\
$2.25$\,<\,$z$\,<\,$2.75$ & 9.5 -- 10 & 0.375 & 0.352 & 0.216 \\
\bottomrule
\end{tabular}
\end{table}

The resulting constraints on the BCLF model parameters in Model~I are shown in Fig.~\ref{fig:param_constraints} for two example stellar mass bins where strong evidence for a decorrelation between H$\alpha$ and $U$-band luminosities may exist. As shown by the ellipses, the cross-correlation between SPHEREx and Rubin/LSST surveys can place useful constraints on our main proxy for bursty star formation, $\sigma_{\alpha U}$, up to $z\sim2.5$, despite the degeneracy between $\sigma_{\alpha U}$ and $\beta$. At redshifts where the constraints are tight enough (e.g., $z\sim1.5$), it is also possible to quantify by how much $\sigma_{\alpha U}$ differs between different mass bins, which serves as another way to probe the strength of star formation burstiness (see Section~\ref{sec:discussion} for further discussion). Table~\ref{tb:fisher_params} summarizes the constraints on the three BCLF model parameters in terms of the fractional uncertainties derived from the diagonal of the inverse of the Fisher matrix in each redshift and mass bin. 

\begin{figure}
    \centering
	\includegraphics[width=\columnwidth]{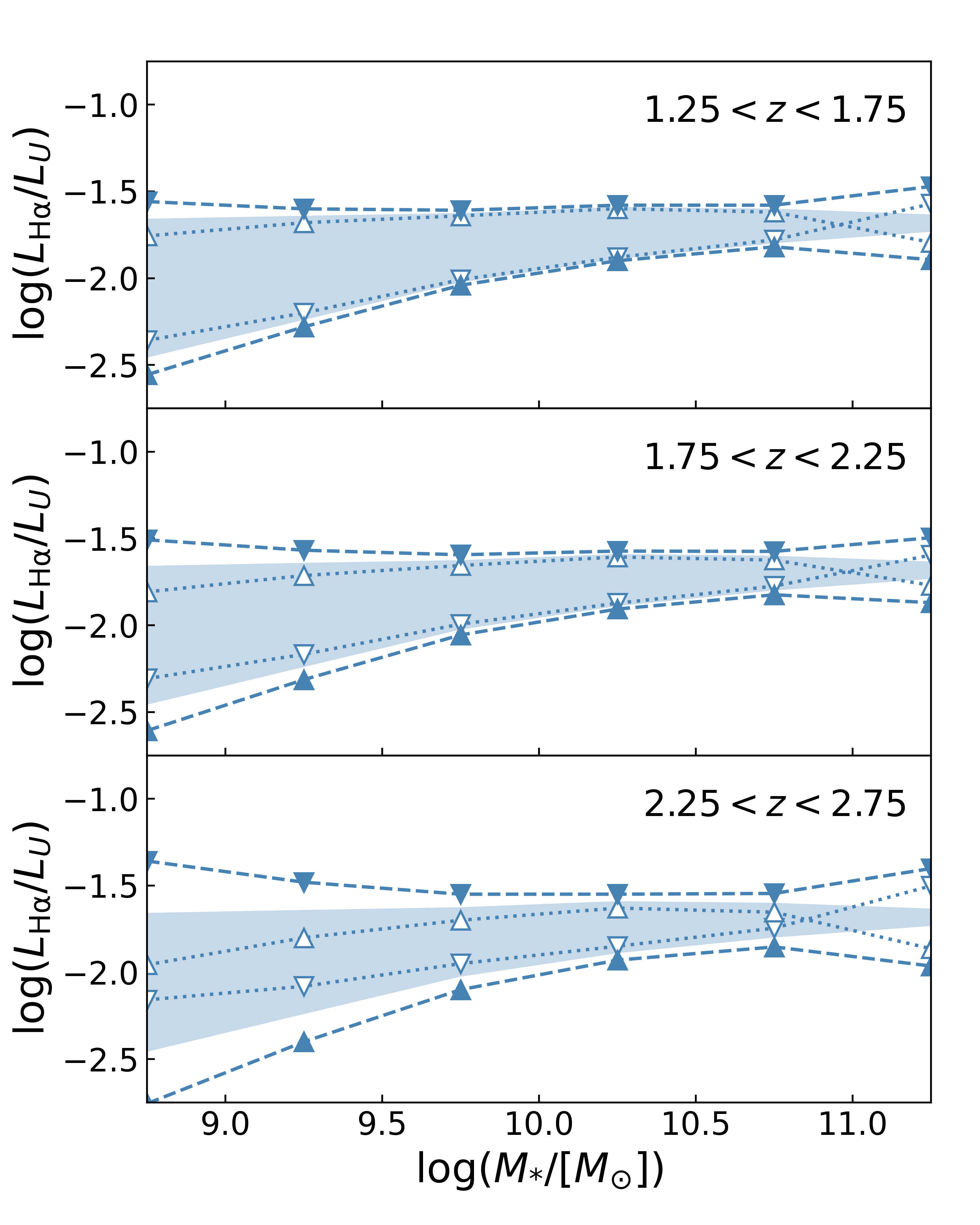}
    \caption{Constraints on $\log(L_\mathrm{H\alpha}/L_{U})$ as a function of $M_{*}$ in the three redshift bins expected from Model~I (shaded band) and the Fisher matrix analysis. Marginalized $\pm 1\sigma$ bounds on the \textit{width} of the $\log(L_\mathrm{H\alpha}/L_{U})$ distribution are indicated by the outer, dashed curves with filled triangles ($1\sigma$ upper bound) and the inner, dotted curves with empty triangles ($1\sigma$ lower bound), respectively. Note that the dotted (lower-bound) curves cross at the high-mass end as a result of increased uncertainties.}
    \label{fig:ratio_constraints}
\end{figure}

Applying the Fisher matrix formalism to all mass and redshift bins and extracting the variance on $\sigma_{\alpha U}$, we derive ultimately the constraints on the $\log(L_\mathrm{H\alpha}/L_{U})$--$M_{*}$ relation available from the EBL--galaxy cross-correlation using SPHEREx and Rubin/LSST data, which can be readily compared with observations of individual galaxies. The resulting constraints are illustrated in Fig.~\ref{fig:ratio_constraints} in terms of the upper (dashed curves and filled triangles) and lower bounds (dotted curves and empty triangles) on the width of the $\log(L_\mathrm{H\alpha}/L_{U})$--$M_{*}$ distribution. From these constraints, it can be seen that any stellar mass dependence of $\sigma_{\alpha U}$ resulting from changes in the SFR variability may be tested by the cross-correlation analysis up to $z\sim2$, beyond which data from SPHEREx and Rubin/LSST can not provide sufficient constraining power.

\section{Discussion} \label{sec:discussion}

By cross-correlating EBL and galaxy surveys to be conducted by SPHEREx and Rubin/LSST as an example, we have so far demonstrated that statistical constraints on the BCLF of $L_\mathrm{H\alpha}$ and $L_\mathrm{UV}$ may be obtained at high significance and used to probe bursty star formation across a wide range of galaxy mass and redshift. Next, we supplement the presented analysis with a semi-quantitative discussion of the caveats, limitations, and implications of the method explored in this work. In particular, we focus on ways to identify and reduce the potential ambiguity from dust attenuation, and compare the statistical approach presented in this paper with the characterization of SFR indicators like $L_\mathrm{H\alpha}$ and $L_\mathrm{UV}$ for samples of individual galaxies.   

\subsection{Ambiguity associated with dust attenuation}

For individual galaxies, both $L_\mathrm{H\alpha}$ and $L_{U}$ are subject to non-negligible dust attenuation, but the amount of attenuation can vary substantially and with different time dependence for H$\alpha$ and UV continuum emission from galaxy to galaxy, as a result of the different sites and mechanisms these photons are created in galaxies (the UV continuum can be much more extended than H$\alpha$ emission produced in star-forming regions). Consequently, part of the observed scatter $\sigma_{\alpha U}$ may actually be associated with variations of the level of dust attenuation rather than star formation burstiness for a given galaxy sample \citep{Reddy_2015}. For the analysis presented, we do not consider the effect of dust on the BCLF. We do, however, take into account dust attenuation in estimating the detectability of various cross-correlation signals. While methods have been proposed to apply appropriate dust corrections for accurate comparison of H$\alpha$ and UV SFR indicators \cite[see e.g.,][for an example method based on energy balance]{Weisz_2012}, they do not directly apply to the statistical approach considered in this paper. Here, through a similar cross-correlation analysis to estimate the Balmer decrement ($L_\mathrm{H\alpha}/L_\mathrm{H\beta}$) variations, we discuss a possible way to reduce the ambiguity associated with unknown dust attenuation variations in the interpretation of results like those shown in Section~\ref{sec:results}. 

The attenuation $A_{\lambda} = k_{\lambda} E(B-V)$ and the Balmer decrement are related by \citep{Dominguez_2013}
\begin{equation}
A_\mathrm{FUV} = \mathcal{C} \log \left( \frac{L_\mathrm{H\alpha}/L_\mathrm{H\beta}}{2.86} \right) , 
\end{equation}
where the coefficient $C = 2.5 k_\mathrm{FUV} / (k_\mathrm{H\beta} - k_\mathrm{H\alpha}) = 19.6$ and $L_\mathrm{H\alpha}/L_\mathrm{H\beta}=2.86$ is the intrinsic Balmer decrement that remains roughly constant for typical star-forming galaxies. Assuming perfectly correlated scatters in dust-attenuated $L_\mathrm{H\alpha}$ and $L_{U}$ induced by a scatter in $A_\mathrm{FUV}$ as defined in equation~(\ref{eq:a_fuv}), we find that a scatter of about 4 in $A_\mathrm{FUV}$, corresponding to a 0.2\,dex scatter in the Balmer decrement $\log(L_\mathrm{H\alpha}/L_\mathrm{H\beta})$, results in a 0.3\,dex scatter in $\log(L_\mathrm{H\alpha}/L_{U})$ comparable to what one might expect from a strongly time-variable SFH. Therefore, to see whether or not an observed scatter in $\log(L_\mathrm{H\alpha}/L_{U})$ can be explained entirely by variations in the dust attenuation, we can use the cross-correlation between H$\alpha$ and H$\beta$ to constrain the scatter $\sigma_\mathrm{BD}$ in $\log(L_\mathrm{H\alpha}/L_\mathrm{H\beta})$. Since $L_\mathrm{H\alpha}$ and $L_\mathrm{H\beta}$ are almost strictly proportional to each other, their cross-correlation coefficient is simply related to $\sigma_\mathrm{BD}$ as $\ln(r^\mathrm{g}_\mathrm{H\alpha \times H\beta, P}) = -\sigma^2_\mathrm{BD}/2$, which implies a more than 10\% decorrelation for a Balmer decrement scatter of $\sigma_\mathrm{BD}=0.2$\,dex. 

At $z\sim1.5$, for example, by performing a detectability analysis for the H$\alpha$--H$\beta$ cross-correlation similar to that shown in Fig.~\ref{fig:b_snr} for the case of H$\alpha$ and $U$-band luminosities, we expect $r^\mathrm{g}_\mathrm{H\alpha \times H\beta, P}$ to be detected at $\mathrm{S/N} \gtrsim 40$ (after redshift binning, see Section~\ref{sec:results:cross-correlations}) by cross-correlating SPHEREx and Rubin/LSST surveys in all stellar mass bins except the least massive one, which is somewhat below the expected mass completeness limit of the Rubin/LSST galaxy survey anyway. Such a high S/N should allow us to reliably test whether or not a notable decorrelation, e.g., $r^\mathrm{g}_\mathrm{H\alpha \times H\beta, P}<0.9$, between $L_\mathrm{H\alpha}$ and $L_\mathrm{H\beta}$ exists as a sign of large variations in the Balmer decrement. This can be compared in turn with level of dust attenuation variations required to fully account for the measured scatter $\sigma_{\alpha U}$ in the H$\alpha$--UV BCLF. 

Finally, it is also noteworthy that the scatter from dust attenuation variations will likely increase with stellar mass, since the massive galaxies tend to be more dust-rich. Therefore, the expected trend with stellar mass is opposite to that of the burstiness, which may also help clarify the ambiguity associated with dust attenuation. 

\subsection{Limitations and implications of the presented method}

Despite its great potential for constraining bursty star formation using forthcoming cosmological survey data sets, the presented framework based on the EBL--galaxy cross-correlation has a few noteworthy limitations due to either simplified assumptions or the methodology itself. First, while being motivated by observations, a rather simplistic description of the BCLF is adopted in this proof-of-concept study. Potentially more self-consistent and physically-grounded models can be constructed from the combination of analytic arguments and results from detailed galaxy simulations, in order to better connect burstiness observables such as $\log(L_\mathrm{H\alpha}/L_{U})$ to realistic representations of the time-variable SFHs. Meanwhile, in the presented analysis we have focused almost entirely on constraining the scatter in the H$\alpha$--UV BCLF, whereas any trend between the mean value and $M_{*}$ may be an additional way to probe bursty star formation. 

Perhaps more importantly, we emphasize the pros and cons of the presented method when compared to observations of individual galaxies with instruments such as \textit{JWST}, which will remain the mainstream approach for studying bursty star formation in the foreseeable future. Because of the requirement on a reference galaxy catalog with not only sufficient depth but also a large sky coverage (nearly 2$\pi$ in the presented case study), it is impossible for the cross-correlation analysis to reach a comparable mass limit to the galaxy observations with \textit{JWST}, which can obtain mass-complete samples of galaxies with more than 100 times lower stellar masses \cite[see e.g.,][]{Bagley_2023}. The most important advantage of the cross-correlation method is the access to a huge sample size of galaxies that probe a much larger cosmic volume than typical pencil-beam-like surveys with \textit{JWST}. At $z=2$, for example, the cumulative number densities of galaxies with stellar mass $M_{*}>10^7\,M_{\odot}$ and $M_{*}>10^9\,M_{\odot}$ only differ by a factor of 10 given the shape of the stellar mass function, whereas photometric galaxy surveys with e.g., Rubin/LSST and Roman typical probe cosmic volumes 1,000 to 10,000 times larger than \textit{JWST} programs. The huge statistical sample of galaxies available for cross-correlation analysis makes it suitable for investigating impacts of e.g., selection bias, cosmic variance, environmental dependence on the interpretation of bursty star formation, in addition to simply providing constraints based on large number statistics that can be cross-checked with direct observations of individual galaxies. 

Notably, the EBL--galaxy cross-correlation as presented already has some interesting implications for better understanding bursty star formation and its role in the process of galaxy formation. With constraints on the H$\alpha$-UV BCLF in different stellar mass and redshift bins, it is possible to identify and distinguish trends predicted by rival galaxy formation and evolution theories involving different assumptions/treatments of the physics of e.g., star formation and stellar feedback. This is particularly of interest during $1 \lesssim z \lesssim 2$ when simulations predict the transition from bursty to steady star formation to happen in many Milky Way progenitor galaxies, which also correlates with the vertical disk settling process \cite[e.g.,][]{Stern_2021, Yu_2022, Gurvich_2023, Hopkins_2023} as a key milestone in the galaxy formation history. Meanwhile, the expected constraints on the scatter of $\log(L_\mathrm{H\alpha}/L_{U})$, and thereby the corresponding level of SFR variability, may also shed light on the possible connection between bursty star formation and the cusp-core transformation of the dark matter halo profiles \cite[e.g.,][]{PontzenGovernato_2012, Teyssier_2013, Chan_2015, Onorbe_2015}. Thanks to the huge sample size available for analyses of multiple sub-samples, the EBL-galaxy cross-correlation can test if the level of stellar feedback and SFR variability required for modulating the dark matter profiles is satisfied in different mass and redshift regimes and cosmological environments. 

\section{Conclusions} \label{sec:conclusions}

We have presented a simple, semi-empirical framework to study the possibility of probing bursty star formation in galaxies at $1 \lesssim z \lesssim 3$ using the cross-correlation between data sets from EBL and galaxy redshift surveys to be available in the next decade. By constructing an observationally-motivated toy model for the BCLF of H$\alpha$ and $U$-band continuum luminosities, two commonly-used SFR indicators probing the recent star formation history on different timescales, we demonstrate how useful constraints on the BCLF can be obtained from Fourier-space analysis of the EBL--galaxy cross-correlation signals in the Poisson-noise limit. Taking the synergy between SPHEREx and Rubin/LSST surveys as an example, we forecast the detectability of key observables derived from the summary statistics, in particular the correlation coefficients, and showcase the expected constraints on the parameter space from these forthcoming data sets. 

Our analysis suggests that useful constraints on the mass and redshift evolution of the BCLF as a key measure of the time variability of the SFH can be placed by the EBL--galaxy cross-correlation, in a complementary manner to traditional methods based on observations of individual galaxies. A similar approach may also be applied to the same data set to investigate the potential ambiguity that can be caused by dust attenuation. In summary, the presented framework provides a novel way to probe bursty star formation and the related physics in high-redshift galaxies using cosmological data sets. Constraints from the EBL--galaxy cross-correlation will be useful additions to deep observations of individual galaxies to be conducted by e.g., \textit{JWST}, thanks to the much greater sample size accessible.

\section*{Acknowledgements}

We thank the anonymous reviewer for their comments, as well as Tzu-Ching Chang and Jordan Mirocha for helpful conversations. G.S. was supported by a CIERA Postdoctoral Fellowship. CAFG was supported by NSF through grants AST-2108230  and CAREER award AST-1652522; by NASA through grants 17-ATP17-0067 and 21-ATP21-0036; by STScI through grant HST-GO-16730.016-A; by CXO through grant TM2-23005X; and by the Research Corporation for Science Advancement through a Cottrell Scholar Award. 

%%%%%%%%%%%%%%%%%%%%%%%%%%%%%%%%%%%%%%%%%%%%%%%%%%
\section*{Data Availability}

The data associated with results presented this article can be shared on reasonable request to the corresponding author.



\bibliographystyle{mnras}
\bibliography{bursty} % if your bibtex file is called example.bib

\begin{thebibliography}{}
\makeatletter
\relax
\def\mn@urlcharsother{\let\do\@makeother \do\$\do\&\do\#\do\^\do\_\do\%\do\~}
\def\mn@doi{\begingroup\mn@urlcharsother \@ifnextchar [ {\mn@doi@}
  {\mn@doi@[]}}
\def\mn@doi@[#1]#2{\def\@tempa{#1}\ifx\@tempa\@empty \href
  {http://dx.doi.org/#2} {doi:#2}\else \href {http://dx.doi.org/#2} {#1}\fi
  \endgroup}
\def\mn@eprint#1#2{\mn@eprint@#1:#2::\@nil}
\def\mn@eprint@arXiv#1{\href {http://arxiv.org/abs/#1} {{\tt arXiv:#1}}}
\def\mn@eprint@dblp#1{\href {http://dblp.uni-trier.de/rec/bibtex/#1.xml}
  {dblp:#1}}
\def\mn@eprint@#1:#2:#3:#4\@nil{\def\@tempa {#1}\def\@tempb {#2}\def\@tempc
  {#3}\ifx \@tempc \@empty \let \@tempc \@tempb \let \@tempb \@tempa \fi \ifx
  \@tempb \@empty \def\@tempb {arXiv}\fi \@ifundefined
  {mn@eprint@\@tempb}{\@tempb:\@tempc}{\expandafter \expandafter \csname
  mn@eprint@\@tempb\endcsname \expandafter{\@tempc}}}

\bibitem[\protect\citeauthoryear{{Bagley} et~al.,}{{Bagley}
  et~al.}{2023}]{Bagley_2023}
{Bagley} M.~B.,  et~al., 2023, \mn@doi [arXiv e-prints]
  {10.48550/arXiv.2302.05466}, \href
  {https://ui.adsabs.harvard.edu/abs/2023arXiv230205466B} {p. arXiv:2302.05466}

\bibitem[\protect\citeauthoryear{{Behroozi}, {Wechsler}, {Hearin}  \&
  {Conroy}}{{Behroozi} et~al.}{2019}]{Behroozi_2019}
{Behroozi} P.,  {Wechsler} R.~H.,  {Hearin} A.~P.,   {Conroy} C.,  2019,
  \mn@doi [\mnras] {10.1093/mnras/stz1182}, \href
  {https://ui.adsabs.harvard.edu/abs/2019MNRAS.488.3143B} {488, 3143}

\bibitem[\protect\citeauthoryear{{Calzetti}, {Armus}, {Bohlin}, {Kinney},
  {Koornneef}  \& {Storchi-Bergmann}}{{Calzetti} et~al.}{2000}]{Calzetti_2000}
{Calzetti} D.,  {Armus} L.,  {Bohlin} R.~C.,  {Kinney} A.~L.,  {Koornneef} J.,
   {Storchi-Bergmann} T.,  2000, \mn@doi [\apj] {10.1086/308692}, \href
  {https://ui.adsabs.harvard.edu/abs/2000ApJ...533..682C} {533, 682}

\bibitem[\protect\citeauthoryear{{Caplar} \& {Tacchella}}{{Caplar} \&
  {Tacchella}}{2019}]{Caplar_2019}
{Caplar} N.,  {Tacchella} S.,  2019, \mn@doi [\mnras] {10.1093/mnras/stz1449},
  \href {https://ui.adsabs.harvard.edu/abs/2019MNRAS.487.3845C} {487, 3845}

\bibitem[\protect\citeauthoryear{{Chan}, {Kere{\v{s}}}, {O{\~n}orbe},
  {Hopkins}, {Muratov}, {Faucher-Gigu{\`e}re}  \& {Quataert}}{{Chan}
  et~al.}{2015}]{Chan_2015}
{Chan} T.~K.,  {Kere{\v{s}}} D.,  {O{\~n}orbe} J.,  {Hopkins} P.~F.,  {Muratov}
  A.~L.,  {Faucher-Gigu{\`e}re} C.~A.,   {Quataert} E.,  2015, \mn@doi [\mnras]
  {10.1093/mnras/stv2165}, \href
  {https://ui.adsabs.harvard.edu/abs/2015MNRAS.454.2981C} {454, 2981}

\bibitem[\protect\citeauthoryear{{Cheng} \& {Bock}}{{Cheng} \&
  {Bock}}{2022}]{CB_2022}
{Cheng} Y.-T.,  {Bock} J.~J.,  2022, \mn@doi [\apj] {10.3847/1538-4357/ac9a51},
  \href {https://ui.adsabs.harvard.edu/abs/2022ApJ...940..115C} {940, 115}

\bibitem[\protect\citeauthoryear{{Cheng} \& {Chang}}{{Cheng} \&
  {Chang}}{2022}]{CC_2022}
{Cheng} Y.-T.,  {Chang} T.-C.,  2022, \mn@doi [\apj]
  {10.3847/1538-4357/ac3aee}, \href
  {https://ui.adsabs.harvard.edu/abs/2022ApJ...925..136C} {925, 136}

\bibitem[\protect\citeauthoryear{{Crill} et~al.,}{{Crill}
  et~al.}{2020}]{Crill_2020}
{Crill} B.~P.,  et~al., 2020, in Society of Photo-Optical Instrumentation
  Engineers (SPIE) Conference Series. p. 114430I, \mn@doi{10.1117/12.2567224}

\bibitem[\protect\citeauthoryear{{Dom{\'\i}nguez} et~al.,}{{Dom{\'\i}nguez}
  et~al.}{2013}]{Dominguez_2013}
{Dom{\'\i}nguez} A.,  et~al., 2013, \mn@doi [\apj]
  {10.1088/0004-637X/763/2/145}, \href
  {https://ui.adsabs.harvard.edu/abs/2013ApJ...763..145D} {763, 145}

\bibitem[\protect\citeauthoryear{{Dom{\'\i}nguez}, {Siana}, {Brooks},
  {Christensen}, {Bruzual}, {Stark}  \& {Alavi}}{{Dom{\'\i}nguez}
  et~al.}{2015}]{Dominguez_2015}
{Dom{\'\i}nguez} A.,  {Siana} B.,  {Brooks} A.~M.,  {Christensen} C.~R.,
  {Bruzual} G.,  {Stark} D.~P.,   {Alavi} A.,  2015, \mn@doi [\mnras]
  {10.1093/mnras/stv1001}, \href
  {https://ui.adsabs.harvard.edu/abs/2015MNRAS.451..839D} {451, 839}

\bibitem[\protect\citeauthoryear{{Dor{\'e}} et~al.,}{{Dor{\'e}}
  et~al.}{2014}]{Dore_2014}
{Dor{\'e}} O.,  et~al., 2014, \mn@doi [arXiv e-prints]
  {10.48550/arXiv.1412.4872}, \href
  {https://ui.adsabs.harvard.edu/abs/2014arXiv1412.4872D} {p. arXiv:1412.4872}

\bibitem[\protect\citeauthoryear{{Emami}, {Siana}, {Weisz}, {Johnson}, {Ma}  \&
  {El-Badry}}{{Emami} et~al.}{2019}]{Emami_2019}
{Emami} N.,  {Siana} B.,  {Weisz} D.~R.,  {Johnson} B.~D.,  {Ma} X.,
  {El-Badry} K.,  2019, \mn@doi [\apj] {10.3847/1538-4357/ab211a}, \href
  {https://ui.adsabs.harvard.edu/abs/2019ApJ...881...71E} {881, 71}

\bibitem[\protect\citeauthoryear{{Emami} et~al.,}{{Emami}
  et~al.}{2021}]{Emami_2021}
{Emami} N.,  et~al., 2021, \mn@doi [\apj] {10.3847/1538-4357/ac1f8d}, \href
  {https://ui.adsabs.harvard.edu/abs/2021ApJ...922..217E} {922, 217}

\bibitem[\protect\citeauthoryear{{Faisst}, {Capak}, {Emami}, {Tacchella}  \&
  {Larson}}{{Faisst} et~al.}{2019}]{Faisst_2019}
{Faisst} A.~L.,  {Capak} P.~L.,  {Emami} N.,  {Tacchella} S.,   {Larson} K.~L.,
   2019, \mn@doi [\apj] {10.3847/1538-4357/ab425b}, \href
  {https://ui.adsabs.harvard.edu/abs/2019ApJ...884..133F} {884, 133}

\bibitem[\protect\citeauthoryear{{Fakhouri}, {Ma}  \&
  {Boylan-Kolchin}}{{Fakhouri} et~al.}{2010}]{Fakhouri_2010}
{Fakhouri} O.,  {Ma} C.-P.,   {Boylan-Kolchin} M.,  2010, \mn@doi [\mnras]
  {10.1111/j.1365-2966.2010.16859.x}, \href
  {https://ui.adsabs.harvard.edu/abs/2010MNRAS.406.2267F} {406, 2267}

\bibitem[\protect\citeauthoryear{{Faucher-Gigu{\`e}re}}{{Faucher-Gigu{\`e}re}}{2018}]{CAFG_2018}
{Faucher-Gigu{\`e}re} C.-A.,  2018, \mn@doi [\mnras] {10.1093/mnras/stx2595},
  \href {https://ui.adsabs.harvard.edu/abs/2018MNRAS.473.3717F} {473, 3717}

\bibitem[\protect\citeauthoryear{{Finke}, {Razzaque}  \& {Dermer}}{{Finke}
  et~al.}{2010}]{Finke_2010}
{Finke} J.~D.,  {Razzaque} S.,   {Dermer} C.~D.,  2010, \mn@doi [\apj]
  {10.1088/0004-637X/712/1/238}, \href
  {https://ui.adsabs.harvard.edu/abs/2010ApJ...712..238F} {712, 238}

\bibitem[\protect\citeauthoryear{{Finke}, {Ajello}, {Dom{\'\i}nguez}, {Desai},
  {Hartmann}, {Paliya}  \& {Saldana-Lopez}}{{Finke} et~al.}{2022}]{Finke_2022}
{Finke} J.~D.,  {Ajello} M.,  {Dom{\'\i}nguez} A.,  {Desai} A.,  {Hartmann}
  D.~H.,  {Paliya} V.~S.,   {Saldana-Lopez} A.,  2022, \mn@doi [\apj]
  {10.3847/1538-4357/ac9843}, \href
  {https://ui.adsabs.harvard.edu/abs/2022ApJ...941...33F} {941, 33}

\bibitem[\protect\citeauthoryear{{Flores Vel{\'a}zquez} et~al.,}{{Flores
  Vel{\'a}zquez} et~al.}{2021}]{JFV_2021}
{Flores Vel{\'a}zquez} J.~A.,  et~al., 2021, \mn@doi [\mnras]
  {10.1093/mnras/staa3893}, \href
  {https://ui.adsabs.harvard.edu/abs/2021MNRAS.501.4812F} {501, 4812}

\bibitem[\protect\citeauthoryear{{Furlanetto} \& {Mirocha}}{{Furlanetto} \&
  {Mirocha}}{2022}]{FM_2022}
{Furlanetto} S.~R.,  {Mirocha} J.,  2022, \mn@doi [\mnras]
  {10.1093/mnras/stac310}, \href
  {https://ui.adsabs.harvard.edu/abs/2022MNRAS.511.3895F} {511, 3895}

\bibitem[\protect\citeauthoryear{{Gong}, {Cooray}, {Silva}, {Zemcov}, {Feng},
  {Santos}, {Dore}  \& {Chen}}{{Gong} et~al.}{2017}]{Gong_2017}
{Gong} Y.,  {Cooray} A.,  {Silva} M.~B.,  {Zemcov} M.,  {Feng} C.,  {Santos}
  M.~G.,  {Dore} O.,   {Chen} X.,  2017, \mn@doi [\apj]
  {10.3847/1538-4357/835/2/273}, \href
  {https://ui.adsabs.harvard.edu/abs/2017ApJ...835..273G} {835, 273}

\bibitem[\protect\citeauthoryear{{Graham}, {Connolly}, {Ivezi{\'c}}, {Schmidt},
  {Jones}, {Juri{\'c}}, {Daniel}  \& {Yoachim}}{{Graham}
  et~al.}{2018}]{Graham_2018}
{Graham} M.~L.,  {Connolly} A.~J.,  {Ivezi{\'c}} {\v{Z}}.,  {Schmidt} S.~J.,
  {Jones} R.~L.,  {Juri{\'c}} M.,  {Daniel} S.~F.,   {Yoachim} P.,  2018,
  \mn@doi [\aj] {10.3847/1538-3881/aa99d4}, \href
  {https://ui.adsabs.harvard.edu/abs/2018AJ....155....1G} {155, 1}

\bibitem[\protect\citeauthoryear{{Graham} et~al.,}{{Graham}
  et~al.}{2020}]{Graham_2020}
{Graham} M.~L.,  et~al., 2020, \mn@doi [\aj] {10.3847/1538-3881/ab8a43}, \href
  {https://ui.adsabs.harvard.edu/abs/2020AJ....159..258G} {159, 258}

\bibitem[\protect\citeauthoryear{{Gurvich} et~al.,}{{Gurvich}
  et~al.}{2023}]{Gurvich_2023}
{Gurvich} A.~B.,  et~al., 2023, \mn@doi [\mnras] {10.1093/mnras/stac3712},
  \href {https://ui.adsabs.harvard.edu/abs/2023MNRAS.519.2598G} {519, 2598}

\bibitem[\protect\citeauthoryear{{Hopkins} et~al.,}{{Hopkins}
  et~al.}{2023}]{Hopkins_2023}
{Hopkins} P.~F.,  et~al., 2023, \mn@doi [arXiv e-prints]
  {10.48550/arXiv.2301.08263}, \href
  {https://ui.adsabs.harvard.edu/abs/2023arXiv230108263H} {p. arXiv:2301.08263}

\bibitem[\protect\citeauthoryear{{Iyer} et~al.,}{{Iyer}
  et~al.}{2020}]{Iyer_2020}
{Iyer} K.~G.,  et~al., 2020, \mn@doi [\mnras] {10.1093/mnras/staa2150}, \href
  {https://ui.adsabs.harvard.edu/abs/2020MNRAS.498..430I} {498, 430}

\bibitem[\protect\citeauthoryear{{Kayo}, {Takada}  \& {Jain}}{{Kayo}
  et~al.}{2013}]{Kayo_2013}
{Kayo} I.,  {Takada} M.,   {Jain} B.,  2013, \mn@doi [\mnras]
  {10.1093/mnras/sts340}, \href
  {https://ui.adsabs.harvard.edu/abs/2013MNRAS.429..344K} {429, 344}

\bibitem[\protect\citeauthoryear{{Korngut} et~al.,}{{Korngut}
  et~al.}{2018}]{Korngut_2018}
{Korngut} P.~M.,  et~al., 2018, in {Lystrup} M.,  {MacEwen} H.~A.,  {Fazio}
  G.~G.,  {Batalha} N.,  {Siegler} N.,   {Tong} E.~C.,  eds,  Society of
  Photo-Optical Instrumentation Engineers (SPIE) Conference Series Vol. 10698,
  Space Telescopes and Instrumentation 2018: Optical, Infrared, and Millimeter
  Wave. p. 106981U, \mn@doi{10.1117/12.2312860}

\bibitem[\protect\citeauthoryear{{LSST Science Collaboration} et~al.,}{{LSST
  Science Collaboration} et~al.}{2009}]{LSST_2009}
{LSST Science Collaboration} et~al., 2009, \mn@doi [arXiv e-prints]
  {10.48550/arXiv.0912.0201}, \href
  {https://ui.adsabs.harvard.edu/abs/2009arXiv0912.0201L} {p. arXiv:0912.0201}

\bibitem[\protect\citeauthoryear{{Laigle} et~al.,}{{Laigle}
  et~al.}{2016}]{Laigle_2016}
{Laigle} C.,  et~al., 2016, \mn@doi [\apjs] {10.3847/0067-0049/224/2/24}, \href
  {https://ui.adsabs.harvard.edu/abs/2016ApJS..224...24L} {224, 24}

\bibitem[\protect\citeauthoryear{{Leauthaud}, {Singh}, {Luo}, {Ardila},
  {Greco}, {Capak}, {Greene}  \& {Mayer}}{{Leauthaud}
  et~al.}{2020}]{Leauthaud_2020}
{Leauthaud} A.,  {Singh} S.,  {Luo} Y.,  {Ardila} F.,  {Greco} J.~P.,  {Capak}
  P.,  {Greene} J.~E.,   {Mayer} L.,  2020, \mn@doi [Physics of the Dark
  Universe] {10.1016/j.dark.2020.100719}, \href
  {https://ui.adsabs.harvard.edu/abs/2020PDU....3000719L} {30, 100719}

\bibitem[\protect\citeauthoryear{{McBride}, {Fakhouri}  \& {Ma}}{{McBride}
  et~al.}{2009}]{McBride_2009}
{McBride} J.,  {Fakhouri} O.,   {Ma} C.-P.,  2009, \mn@doi [\mnras]
  {10.1111/j.1365-2966.2009.15329.x}, \href
  {https://ui.adsabs.harvard.edu/abs/2009MNRAS.398.1858M} {398, 1858}

\bibitem[\protect\citeauthoryear{{McLure} et~al.,}{{McLure}
  et~al.}{2018}]{McLure_2018}
{McLure} R.~J.,  et~al., 2018, \mn@doi [\mnras] {10.1093/mnras/sty522}, \href
  {https://ui.adsabs.harvard.edu/abs/2018MNRAS.476.3991M} {476, 3991}

\bibitem[\protect\citeauthoryear{{Mehta} et~al.,}{{Mehta}
  et~al.}{2015}]{Mehta_2015}
{Mehta} V.,  et~al., 2015, \mn@doi [\apj] {10.1088/0004-637X/811/2/141}, \href
  {https://ui.adsabs.harvard.edu/abs/2015ApJ...811..141M} {811, 141}

\bibitem[\protect\citeauthoryear{{Mirocha}, {Liu}  \& {La Plante}}{{Mirocha}
  et~al.}{2022}]{Mirocha_2022}
{Mirocha} J.,  {Liu} A.,   {La Plante} P.,  2022, \mn@doi [\mnras]
  {10.1093/mnras/stac2530}, \href
  {https://ui.adsabs.harvard.edu/abs/2022MNRAS.516.4123M} {516, 4123}

\bibitem[\protect\citeauthoryear{{More}, {van den Bosch}, {Cacciato}, {Mo},
  {Yang}  \& {Li}}{{More} et~al.}{2009}]{More_2009}
{More} S.,  {van den Bosch} F.~C.,  {Cacciato} M.,  {Mo} H.~J.,  {Yang} X.,
  {Li} R.,  2009, \mn@doi [\mnras] {10.1111/j.1365-2966.2008.14095.x}, \href
  {https://ui.adsabs.harvard.edu/abs/2009MNRAS.392..801M} {392, 801}

\bibitem[\protect\citeauthoryear{{Moutard}, {Sawicki}, {Arnouts}, {Golob},
  {Coupon}, {Ilbert}, {Yang}  \& {Gwyn}}{{Moutard} et~al.}{2020}]{Moutard_2020}
{Moutard} T.,  {Sawicki} M.,  {Arnouts} S.,  {Golob} A.,  {Coupon} J.,
  {Ilbert} O.,  {Yang} X.,   {Gwyn} S.,  2020, \mn@doi [\mnras]
  {10.1093/mnras/staa706}, \href
  {https://ui.adsabs.harvard.edu/abs/2020MNRAS.494.1894M} {494, 1894}

\bibitem[\protect\citeauthoryear{{O{\~n}orbe}, {Boylan-Kolchin}, {Bullock},
  {Hopkins}, {Kere{\v{s}}}, {Faucher-Gigu{\`e}re}, {Quataert}  \&
  {Murray}}{{O{\~n}orbe} et~al.}{2015}]{Onorbe_2015}
{O{\~n}orbe} J.,  {Boylan-Kolchin} M.,  {Bullock} J.~S.,  {Hopkins} P.~F.,
  {Kere{\v{s}}} D.,  {Faucher-Gigu{\`e}re} C.-A.,  {Quataert} E.,   {Murray}
  N.,  2015, \mn@doi [\mnras] {10.1093/mnras/stv2072}, \href
  {https://ui.adsabs.harvard.edu/abs/2015MNRAS.454.2092O} {454, 2092}

\bibitem[\protect\citeauthoryear{{Orr}, {Fielding}, {Hayward}  \&
  {Burkhart}}{{Orr} et~al.}{2022}]{Orr_2022}
{Orr} M.~E.,  {Fielding} D.~B.,  {Hayward} C.~C.,   {Burkhart} B.,  2022,
  \mn@doi [\apj] {10.3847/1538-4357/ac6c26}, \href
  {https://ui.adsabs.harvard.edu/abs/2022ApJ...932...88O} {932, 88}

\bibitem[\protect\citeauthoryear{{Planck Collaboration} et~al.,}{{Planck
  Collaboration} et~al.}{2016}]{Planck_2016XIII}
{Planck Collaboration} et~al., 2016, \mn@doi [\aap]
  {10.1051/0004-6361/201525830}, \href
  {https://ui.adsabs.harvard.edu/abs/2016A&A...594A..13P} {594, A13}

\bibitem[\protect\citeauthoryear{{Pontzen} \& {Governato}}{{Pontzen} \&
  {Governato}}{2012}]{PontzenGovernato_2012}
{Pontzen} A.,  {Governato} F.,  2012, \mn@doi [\mnras]
  {10.1111/j.1365-2966.2012.20571.x}, \href
  {https://ui.adsabs.harvard.edu/abs/2012MNRAS.421.3464P} {421, 3464}

\bibitem[\protect\citeauthoryear{{Reddy} et~al.,}{{Reddy}
  et~al.}{2015}]{Reddy_2015}
{Reddy} N.~A.,  et~al., 2015, \mn@doi [\apj] {10.1088/0004-637X/806/2/259},
  \href {https://ui.adsabs.harvard.edu/abs/2015ApJ...806..259R} {806, 259}

\bibitem[\protect\citeauthoryear{{Scott}, {Upton Sanderbeck}  \&
  {Bird}}{{Scott} et~al.}{2022}]{Scott_2022}
{Scott} B.~R.,  {Upton Sanderbeck} P.,   {Bird} S.,  2022, \mn@doi [\mnras]
  {10.1093/mnras/stac265}, \href
  {https://ui.adsabs.harvard.edu/abs/2022MNRAS.511.5158S} {511, 5158}

\bibitem[\protect\citeauthoryear{{Sheth}, {Mo}  \& {Tormen}}{{Sheth}
  et~al.}{2001}]{SMT_2001}
{Sheth} R.~K.,  {Mo} H.~J.,   {Tormen} G.,  2001, \mn@doi [\mnras]
  {10.1046/j.1365-8711.2001.04006.x}, \href
  {https://ui.adsabs.harvard.edu/abs/2001MNRAS.323....1S} {323, 1}

\bibitem[\protect\citeauthoryear{{Shuntov} et~al.,}{{Shuntov}
  et~al.}{2022}]{Shuntov_2022}
{Shuntov} M.,  et~al., 2022, \mn@doi [\aap] {10.1051/0004-6361/202243136},
  \href {https://ui.adsabs.harvard.edu/abs/2022A&A...664A..61S} {664, A61}

\bibitem[\protect\citeauthoryear{{Sparre}, {Hayward}, {Feldmann},
  {Faucher-Gigu{\`e}re}, {Muratov}, {Kere{\v{s}}}  \& {Hopkins}}{{Sparre}
  et~al.}{2017}]{Sparre_2017}
{Sparre} M.,  {Hayward} C.~C.,  {Feldmann} R.,  {Faucher-Gigu{\`e}re} C.-A.,
  {Muratov} A.~L.,  {Kere{\v{s}}} D.,   {Hopkins} P.~F.,  2017, \mn@doi
  [\mnras] {10.1093/mnras/stw3011}, \href
  {https://ui.adsabs.harvard.edu/abs/2017MNRAS.466...88S} {466, 88}

\bibitem[\protect\citeauthoryear{{Spergel} et~al.,}{{Spergel}
  et~al.}{2015}]{Spergel_2015}
{Spergel} D.,  et~al., 2015, \mn@doi [arXiv e-prints]
  {10.48550/arXiv.1503.03757}, \href
  {https://ui.adsabs.harvard.edu/abs/2015arXiv150303757S} {p. arXiv:1503.03757}

\bibitem[\protect\citeauthoryear{{Stern} et~al.,}{{Stern}
  et~al.}{2021}]{Stern_2021}
{Stern} J.,  et~al., 2021, \mn@doi [\apj] {10.3847/1538-4357/abd776}, \href
  {https://ui.adsabs.harvard.edu/abs/2021ApJ...911...88S} {911, 88}

\bibitem[\protect\citeauthoryear{{Sun}}{{Sun}}{2022}]{Sun_2022}
{Sun} G.,  2022, \mn@doi [\apjl] {10.3847/2041-8213/ac7138}, \href
  {https://ui.adsabs.harvard.edu/abs/2022ApJ...931L..29S} {931, L29}

\bibitem[\protect\citeauthoryear{{Sun} \& {Furlanetto}}{{Sun} \&
  {Furlanetto}}{2016}]{SF_2016}
{Sun} G.,  {Furlanetto} S.~R.,  2016, \mn@doi [\mnras] {10.1093/mnras/stw980},
  \href {https://ui.adsabs.harvard.edu/abs/2016MNRAS.460..417S} {460, 417}

\bibitem[\protect\citeauthoryear{{Sun}, {Mirocha}, {Mebane}  \&
  {Furlanetto}}{{Sun} et~al.}{2021}]{Sun_2021}
{Sun} G.,  {Mirocha} J.,  {Mebane} R.~H.,   {Furlanetto} S.~R.,  2021, \mn@doi
  [\mnras] {10.1093/mnras/stab2697}, \href
  {https://ui.adsabs.harvard.edu/abs/2021MNRAS.508.1954S} {508, 1954}

\bibitem[\protect\citeauthoryear{{Sun}, {Faucher-Gigu{\`e}re}, {Hayward}  \&
  {Shen}}{{Sun} et~al.}{2023}]{Sun_2023}
{Sun} G.,  {Faucher-Gigu{\`e}re} C.-A.,  {Hayward} C.~C.,   {Shen} X.,  2023,
  \mn@doi [arXiv e-prints] {10.48550/arXiv.2305.02713}, \href
  {https://ui.adsabs.harvard.edu/abs/2023arXiv230502713S} {p. arXiv:2305.02713}

\bibitem[\protect\citeauthoryear{{Teyssier}, {Pontzen}, {Dubois}  \&
  {Read}}{{Teyssier} et~al.}{2013}]{Teyssier_2013}
{Teyssier} R.,  {Pontzen} A.,  {Dubois} Y.,   {Read} J.~I.,  2013, \mn@doi
  [\mnras] {10.1093/mnras/sts563}, \href
  {https://ui.adsabs.harvard.edu/abs/2013MNRAS.429.3068T} {429, 3068}

\bibitem[\protect\citeauthoryear{{Viero}, {Sun}, {Chung}, {Moncelsi}  \&
  {Condon}}{{Viero} et~al.}{2022}]{Viero_2022}
{Viero} M.~P.,  {Sun} G.,  {Chung} D.~T.,  {Moncelsi} L.,   {Condon} S.~S.,
  2022, \mn@doi [\mnras] {10.1093/mnrasl/slac075}, \href
  {https://ui.adsabs.harvard.edu/abs/2022MNRAS.516L..30V} {516, L30}

\bibitem[\protect\citeauthoryear{{Weisz} et~al.,}{{Weisz}
  et~al.}{2012}]{Weisz_2012}
{Weisz} D.~R.,  et~al., 2012, \mn@doi [\apj] {10.1088/0004-637X/744/1/44},
  \href {https://ui.adsabs.harvard.edu/abs/2012ApJ...744...44W} {744, 44}

\bibitem[\protect\citeauthoryear{{Yang}, {Mo}  \& {van den Bosch}}{{Yang}
  et~al.}{2003}]{Yang_2003}
{Yang} X.,  {Mo} H.~J.,   {van den Bosch} F.~C.,  2003, \mn@doi [\mnras]
  {10.1046/j.1365-8711.2003.06254.x}, \href
  {https://ui.adsabs.harvard.edu/abs/2003MNRAS.339.1057Y} {339, 1057}

\bibitem[\protect\citeauthoryear{{Yu} et~al.,}{{Yu} et~al.}{2022}]{Yu_2022}
{Yu} S.,  et~al., 2022, \mn@doi [arXiv e-prints] {10.48550/arXiv.2210.03845},
  \href {https://ui.adsabs.harvard.edu/abs/2022arXiv221003845Y} {p.
  arXiv:2210.03845}

\bibitem[\protect\citeauthoryear{{Zheng} et~al.,}{{Zheng}
  et~al.}{2005}]{Zheng_2005}
{Zheng} Z.,  et~al., 2005, \mn@doi [\apj] {10.1086/466510}, \href
  {https://ui.adsabs.harvard.edu/abs/2005ApJ...633..791Z} {633, 791}

\bibitem[\protect\citeauthoryear{{Zhou} et~al.,}{{Zhou}
  et~al.}{2017}]{Zhou_2017}
{Zhou} Z.,  et~al., 2017, \mn@doi [\apj] {10.3847/1538-4357/835/1/70}, \href
  {https://ui.adsabs.harvard.edu/abs/2017ApJ...835...70Z} {835, 70}

\bibitem[\protect\citeauthoryear{{van den Bosch}, {Jiang}, {Hearin},
  {Campbell}, {Watson}  \& {Padmanabhan}}{{van den Bosch}
  et~al.}{2014}]{vandenBosch_2014}
{van den Bosch} F.~C.,  {Jiang} F.,  {Hearin} A.,  {Campbell} D.,  {Watson} D.,
    {Padmanabhan} N.,  2014, \mn@doi [\mnras] {10.1093/mnras/stu1872}, \href
  {https://ui.adsabs.harvard.edu/abs/2014MNRAS.445.1713V} {445, 1713}

\makeatother
\end{thebibliography}

%\begin{thebibliography}{99}
%\end{thebibliography}

%%%%%%%%%%%%%%%%%%%%%%%%%%%%%%%%%%%%%%%%%%%%%%%%%%

%%%%%%%%%%%%%%%%% APPENDICES %%%%%%%%%%%%%%%%%%%%%

\appendix

\section{Connections to Other Statistics} \label{appendix:taylor}

While $r^\mathrm{g}_\mathrm{\times,P}$ itself already encodes valuable information about the burstiness of star formation, there are other observables that are also informative about the variability of the SFR. Previous studies have suggested that the full PDF, $P(L_\mathrm{H\alpha}/L_{U})$, may be the most reliable probe of bursty star formation. Thus, it might also be useful to inspect basic measures of $P(L_\mathrm{H\alpha}/L_{U})$ like the mean, $\langle L_\mathrm{H\alpha}/L_{U} \rangle$, and the variance, $\mathrm{Var} \left( L_\mathrm{H\alpha}/L_{U} \right)$. Using Taylor expansion to the second order, we can approximate them as
\begin{equation}
\bigg<\frac{L_\mathrm{H\alpha}}{L_{U}}\bigg> = \frac{\langle L_\mathrm{H\alpha} \rangle}{\langle L_{U} \rangle} - \frac{\mathrm{Cov}(L_\mathrm{H\alpha}, L_{U})}{\langle L_{U} \rangle^2} + \frac{\mathrm{Var}(L_{U}) \langle L_\mathrm{H\alpha} \rangle}{\langle L_{U} \rangle^3}
\end{equation}
and
\begin{align}
\mathrm{Var}\left( \frac{L_\mathrm{H\alpha}}{L_{U}} \right) = &\ \bigg[ \frac{\mathrm{Var}(L_\mathrm{H\alpha})}{\langle L_\mathrm{H\alpha} \rangle^2} - \frac{2 \mathrm{Cov}(L_\mathrm{H\alpha}, L_{U})}{\langle L_\mathrm{H\alpha} \rangle \langle L_{U} \rangle} + \frac{\mathrm{Var}(L_{U})}{\langle L_{U} \rangle^2} \bigg] \nonumber \\ & \times \frac{\langle L_\mathrm{H\alpha} \rangle^2}{\langle L_{U} \rangle^2}.
\end{align}
By definition, the covariance and variances can be written as $\mathrm{Cov}(L_\mathrm{H\alpha}, L_{U}) = \langle L_\mathrm{H\alpha} L_{U} \rangle - \langle L_\mathrm{H\alpha} \rangle \langle L_{U} \rangle$, $\mathrm{Var}(L_\mathrm{H\alpha}) = \langle L_\mathrm{H\alpha}^2 \rangle - \langle L_\mathrm{H\alpha} \rangle^2$, and $\mathrm{Var}(L_{U}) = \langle L_{U}^2 \rangle - \langle L_{U} \rangle^2$. As shown in Section~\ref{sec:psbs}, while $\langle \mathcal{O}(L) \rangle$ terms can be readily derived from cross-power spectra of intensity maps and galaxies, we must resort to the UV--UV--galaxy, H$\alpha$--H$\alpha$--galaxy, and H$\alpha$--UV--galaxy cross-bispectra to estimate the $\langle \mathcal{O}(L^2) \rangle$ terms, as in the case of $r^\mathrm{g}_\mathrm{\times,P}$. 

\section{Biases of Galaxies and Intensities of H$\alpha$ and UV Emission} \label{appendix:bias}

In what follows, we specify how we estimate biases of the tracers of interest (galaxies, H$\alpha$ and UV emission). While these bias factors do not directly enter our main analysis, which is limited to the Poisson-noise-dominated regime, they are essential for verifying the range of valid $\ell$'s where our analysis should be performed, without substantial contamination from the clustering signals. Generally, we can express the average bias factor as
\begin{equation}
b_i (z) = \frac{\int d M_* \psi(M_*,z) b(M_*f^{-1}_\mathrm{SHMR}, z) \mathcal{W}_i(M_*, z)}{\int d M_* \psi(M_*,z) \mathcal{W}_i(M_*, z)},
\end{equation}
where $b(M, z)$ is the halo bias \citep{SMT_2001}, $f^{-1}_\mathrm{SHMR}$ is the inverse of the stellar-to-halo mass ratio \citep{Behroozi_2019}, and $\mathcal{W}_i$ is some weighting function that varies for different signals. For the bias factor of galaxies, $\langle b \rangle_\mathrm{g}(z)$, we assume $\mathcal{W}_\mathrm{g}(M_*, z) = \langle N_\mathrm{g} \rangle (M_*f^{-1}_\mathrm{SHMR}, z)$, where $N_\mathrm{g} = N_\mathrm{cen} + N_\mathrm{sat}$ follows the halo occupation distribution (HOD) parameterization in \citet{Zheng_2005}. To estimate the bias factors of H$\alpha$ and UV emission, $b_\mathrm{H\alpha}(z)$ and $b_{U}(z)$, we assume $\mathcal{W}_\mathrm{H\alpha}(M_*, z) = \bar{L}_\mathrm{H\alpha}[\bar{L}_{U}(M_*), z]$ and $\mathcal{W}_{U}(M_*, z) = \bar{L}_{U}(M_*, z)$, respectively. 

%%%%%%%%%%%%%%%%%%%%%%%%%%%%%%%%%%%%%%%%%%%%%%%%%%

% Don't change these lines
\bsp	% typesetting comment
\label{lastpage}
\end{document}